# Modal analysis of electromagnetic resonators: user guide for the MAN program

T. Wu[1,†], D. Arrivault[1], W. Yan[2,3], P. Lalanne[1,*]

[1]LP2N, Institut d'Optique Graduate School, CNRS, Univ. Bordeaux, 33400 Talence, France

[2]Key Laboratory of 3D Micro/Nano Fabrication and Characterization of Zhejiang Province, School of Engineering, Westlake University, 18 Shilongshan Road, Hangzhou 310024, Zhejiang Province, China

[3]Institute of Advanced Technology, Westlake Institute for Advanced Study, 18 Shilongshan Road, Hangzhou 310024, Zhejiang Province, China

\* philippe.lalanne@institutoptique.fr

† wutong1121@gmail.com

## Abstract

All electromagnetic systems, in particular resonators or antennas, have resonances with finite lifetimes. The associated eigenstates, also called quasinormal modes, are essentially non-Hermitian and determine the optical responses of the system. We introduce **MAN** (**M**odal **A**nalysis of **N**anoresonators), a software with many open scripts, which computes and normalizes the quasinormal modes of virtually any electromagnetic resonator, be it composed of dispersive, anisotropic, or non-reciprocal materials. **MAN** reconstructs the scattered field in the basis formed by the quasinormal modes of the resonator and provides a transparent interpretation of the physics.

The software is implemented in MATLAB and has been developed over the past ten years. **MAN** features many toolboxes that illustrate how to use the software for various emblematic computations in low and high frequency regimes. A specific effort has been devoted to interface the solver with the finite-element software COMSOL Multiphysics. However, **MAN** can also be used with other frequency-domain numerical solvers. This article introduces the program and summarizes the relevant theoretical background. **MAN** includes a comprehensive set of classical models and toolboxes that can be downloaded from the web.

## Program Summary

Program title: MAN (Model Analysis of Nanoresonators)

CPC Library link to program files: XXXXX

Licensing provisions: GNU General Public License 3

Developer's repository link: DOI:10.5281/zenodo.7101922

Programming language: Matlab

Nature of problem: Compute and normalize the quasinormal modes of resonators and antennas of any type, be they composed of dispersive, anisotropic, or non-reciprocal materials, or be they operated at high or low frequencies. Use the quasinormal modes to reconstruct the resonator responses in the near field and far field.

Solution method: Two methods are gathered. The quasinormal modes are computed and normalized either by solving the linearized Maxwell equations or searching poles in the complex frequency plane. The reconstruction is then performed analytically with closed form expressions of the excitation coefficients.

**KEYWORDS**: Computational electromagnetic methods, Eigenmode Solver, Microwave Cavities, RF Antennas, Nanoresonators, Microcavities, Gratings, Photonic Crystals, Plasmon, Quasinormal mode.

# 1 Motivation

The interaction of electromagnetic fields with resonators and antennas is driven by the natural resonances of the system. Under excitation by a pulse, the resonances are initially loaded, before releasing their energy by exponentially decaying. The decay is due to absorption or leakage and thus resonators are non-Hermitian systems that do not conserve their energy. This is why their supported resonances are called quasinormal modes (QNMs), therein emphasizing that they are the eigenvectors of a non-Hermitian operator, the Maxwell operator, with complex-valued eigenfrequencies, $\widetilde{\omega} = \Omega + i\Gamma/2$, where $\Gamma$ stands for the decay rate, i.e., the inverse of the mode lifetime $\tau = 1/\Gamma$. The factor $1/2$ accounts for the difference between amplitude and energy decays.

QNMs are formally defined as the time-harmonic solutions to the source-free Maxwell equations ($\exp(i\omega t)$ convention) [1,2]

$$\nabla \times \widetilde{\mathbf{E}} = -i\widetilde{\omega}\boldsymbol{\mu}(\mathbf{r},\widetilde{\omega})\widetilde{\mathbf{H}},$$
$$\nabla \times \widetilde{\mathbf{H}} = i\widetilde{\omega}\boldsymbol{\varepsilon}(\mathbf{r},\widetilde{\omega})\widetilde{\mathbf{E}}, \qquad (1)$$

+ outgoing waves conditions.

Throughout the article, $[\widetilde{\mathbf{E}}(\mathbf{r}), \widetilde{\mathbf{H}}(\mathbf{r})]$ denotes the electric and magnetic fields of *normalized* QNMs. Normalization is central in QNM theory. It is not documented in the present article; for more information, please refer to other sources, e.g. [1,3] and especially Section 4 in [3] that clarifies a long standing debate and compares the four main approaches that have been used in the literature during the last decade with respect to their exactness, domain of validity, and accuracy of their numerical implementation.

In this work, we present **MAN** (**M**odal **A**nalysis of **N**anoresonator), an open-source software that uses the QNM basis for analyzing the response of virtually any resonators or antennas, be they composed of dispersive, anisotropic, or non-reciprocal materials, or operated at high (optical and near-IR waves) or low (RF waves) frequencies. **MAN** is conceived to educatively help the user analyze the physics of electromagnetic resonators towards the identification of the dominant QNMs to speed up the simulation of resonant structures and offer physical transparency in the design.

**MAN** is the result of a collective effort, initiated at the Institut d'Optique in Palaiseau by Jean-Paul Hugonin, Christophe Sauvan and Philippe Lalanne [4,5] and continued in Bordeaux over the last decade [1,3,6-13]. Because the effort dominantly targets nanophotonic applications, the examples and models are provided at optical frequencies, but the approach can be easily extended to the low-frequency domain, as exemplified by the nonreciprocal toolbox which provides an example with millimeter waves.

**MAN** V8 (and following versions) is also much more user-friendly and more general than previous versions. By gathering the two first available freeware, **QNMPole** [4,5] and **QNMEig** [7], which were respectively launched in 2013 and 2018, it provides two efficient, independent, and mathematically-equivalent methods [3] to compute and normalize QNMs. It also provides various toolboxes for reconstructing the response, e.g. the field scattered by the resonator, the extinction and absorption cross-sections, and the Purcell factor, as a superposition of QNMs. A special attention is devoted to the computation of key QNM figures of merit, such as the quality factor and the mode volume [1,12], a complex quantity for non-Hermitian systems that is central to the resonant interaction of light with point-like scatterers or emitters. It also informs the user on important QNM properties, such as their far-field radiation pattern, their multipolar nature, or their brightness for absorptive resonators.

# 2 Underlying principles of the code

**2.1 Overview**

Figure 1 shows the structure of the software.

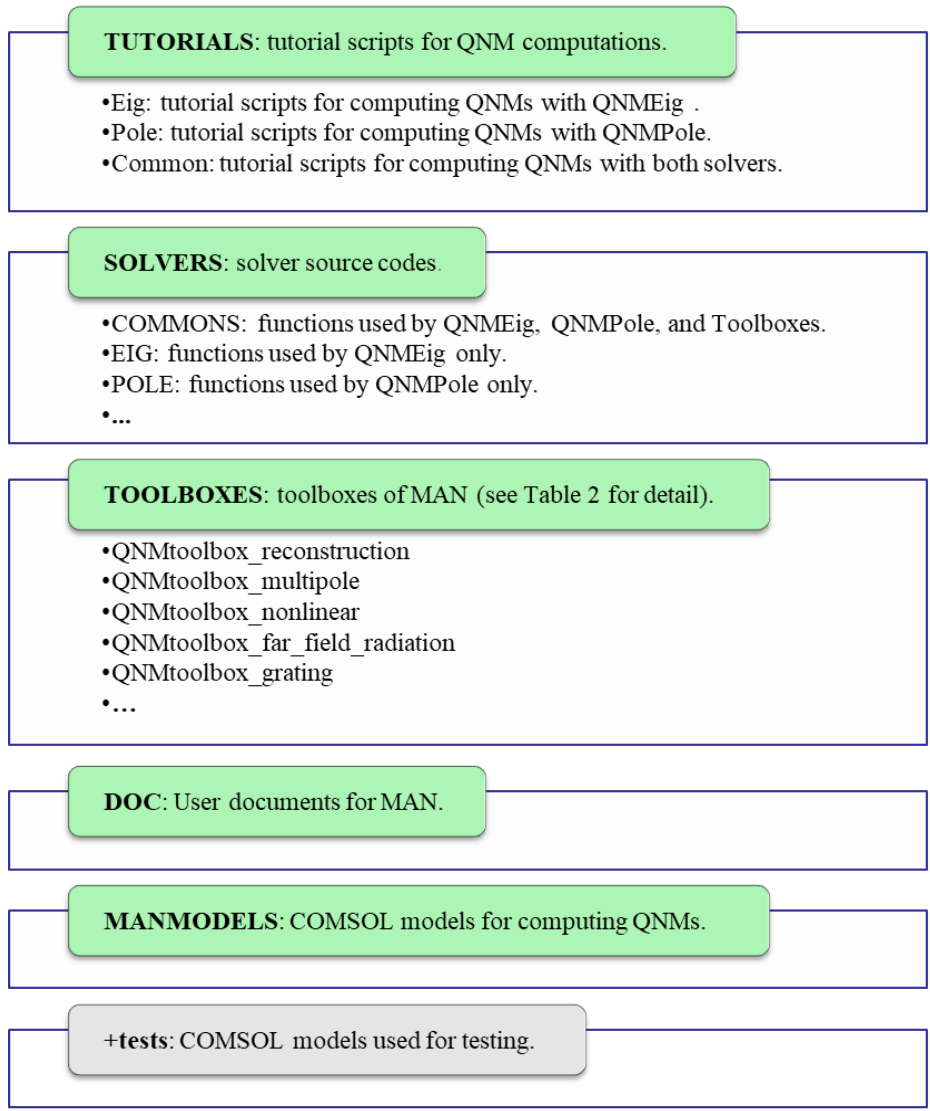

**Fig. 1.** Structure of **MAN**.

The source code is composed of four main folders: *TUTORIALS*, *SOLVERS*, *TOOLBOXES,* and *DOC*.

- *TUTORIALS* folder: The *Common* subfolder of the *TUTORIALS* folder contains the MATLAB scripts Tutorial1D.m, Tutorial2D.m, and Tutorial3D.m. Tutorial1D.m illustrates how one may simply compute, normalize, sort, and save the QNMs computed for a very simple geometry, a 1D Fabry-Perot cavity. It solely relies on MATLAB and serves as a pedagogical introduction to Tutorial2D.m and Tutorial3D.m, which apply to complicated geometries thanks to the solvers **QNMEig** or **QNMPole**.
- *TOOLBOXES* folder: The QNM solvers are valued by an increasing number of toolboxes gathered in the *TOOLBOXES* folder. The toolboxes contain illustrative examples and user guides of emblematic problems in nanophotonics. A detailed introduction to the toolboxes is found in Section 4. The toolboxes are solver dependent; this is quite formal since, with a minor effort from the user, the toolboxes developed for one solver may be adapted for use with the other solver.
- *DOC* folder: The *DOC* folder contains user guides for **QNMEig**, **QNMPole**, and all the toolboxes.
- *SOLVERS* folder: The *SOLVERS* folder contains all the MATLAB functions used by the QNM solvers and companion scripts for QNMs computation, postprocessing, and visualization.

- *MANMODELS* folder: The *MANMODELS* folder is composed of over 20 COMSOL models, including photonic crystal cavities, plasmonic nanospheres, nano-particle-on-mirror (NPoM) geometries… We plan to increase the number of models over time. The models are solver dependent and can be used with either **QNMEig** or **QNMPole**, but not with both solvers. Note that the QNMEig models can be run independently of any MATLAB script in **MAN** if the user goal is just to compute the normalized QNMs.

## 2.2 How to cite?

We kindly ask that you acknowledge the **MAN** package and its authors in any publication/report for which you use it. The preferred citation for **MAN** is the present paper.

In addition, the authors of the toolboxes and the QNM solvers are different from those of the present publication. It would be fair to quote Refs. [5] or [7] if one uses **QNMPole** or **QNMEig**, respectively. For the citation of the toolboxes, please refer to Table 2.

## 2.3 QNM solvers

To cover a broad scope of applications, **MAN** implements two largely different QNM solvers, **QNMEig** [7] and **QNMPole** [5], which compute and normalize QNMs with completely different, albeit mathematically equivalent [3], numerical methods. Both solvers need to be interfaced with a third-party Maxwell solver provided by the user to be operational.

**QNMEig** is dedicated for use with COMSOL Multiphysics and can be viewed as an extension of the eigenmode solver of the COMSOL RF Module. Comparatively, **QNMPole** is much simpler and more general; it is difficult to conceive of a simpler solver. **QNMPole** can be straightforwardly interfaced with any frequency-domain simulation software, be they commercial or in-house.

In **MAN**, **QNMPole** is interfaced with the frequency domain solver of the COMSOL RF Module. However, the interfacing principle can be adapted with minimal effort to any other third-party frequency-domain solvers, including simulation software dedicated to the design and simulation of high-frequency electronic products (antennas, RF or microwave components…), such as the Frequency Domain FEM of Ansys HFSS [14] or the Frequency Domain Solver of CST Studio Suite [15]. Most FDTD packages, such as MEEP [16] or Lumerical FDTD of Ansys [17], incorporates a frequency-domain solver, and can also be easily interfaced with **QNMPole**.

**QNMPole** and **QNMEig** strictly provide the same eigenfrequencies and normalized QNMs if they are both used with the same discretization technique. Their main characteristics are summarized in Table 1.

**Table 1.** Summary of the QNMEig and QNMPole solvers.

| Solver | QNM computation method | Constitutive materials | Companion software needed |
|---|---|---|---|
| **QNMEig** [7] | The RF Module eigenmode solver is used to solve a linearized eigenproblem augmented with auxiliary fields. | Lorentz-Drude: $\varepsilon(\omega)/\varepsilon_\infty = 1 - \Sigma_{n=1}^{N} \omega_{pn}^2/(\omega^2 - \omega_{0n}^2 - i\omega\gamma_n)$. <br> N-pole Lorentz: $\varepsilon(\omega) = \varepsilon_\infty + \Sigma_{n=1}^{N}[A_n/(\omega - \omega_n) - A_n^*/(\omega + \omega_n^*)]$. <br> Constant tensor | • COMSOL Multiphysics RF Module. <br> • COMSOL Multiphysics Weak-form Module. <br> • MATLAB (for postprocessing and data visualization). |
| **QNMPole** [5] | Iterative pole search in the complex frequency plane. | Any permittivity or permeability with a frequency dependence that is analytically known at complex frequencies. | • Any software incorporating a frequency-domain Maxwell solver. <br> • MATLAB (for driving Maxwell solvers to search poles, |

| | | | postprocessing, and data visualization). |
|--|--|--|--|

### 2.3.1 QNMEig

The RF Module eigenmode solver of COMSOL takes advantage of state-of-the art numerical tools, such as the Advanced Krylov Subspace method and the Jacobi-Davidson method [18]. It is a professional tool that is highly effective and accurate. However, it is strictly valid only for QNMs of resonators and antennas made of non-dispersive materials. For dispersive materials, the permittivity (or permeability) depends on the eigenvalue and the Maxwell operator of Eq. (1) defines a non-linear eigenproblem. The RF Module eigenmode solver approximately linearizes the non-linear eigenequation by providing a second-order Taylor expansion of the complex material parameters with respect to some frequency set by the user, the so-called linearization point [18]. COMSOL then computes the eigenvalue by making a few iterations, updating the linearization point at each step. According to our experience, the iteration is convergent only for weakly dispersive materials, such as dielectrics far from their absorption lines.

**QNMEig** uses the RF Module eigenmode solver and thus benefits from its advanced framework, but additionally incorporates auxiliary fields by coupling the built-in RF Module and the Weak-form Module. The auxiliary fields are used to linearize the eigenproblem, but this time, the linearization is exact and a large number (assigned by the user) of QNMs are computed with high accuracy [7], even for highly dispersive materials, such as Drude metals or multipole Drude-Lorentz materials. The QNMs are then normalized by computing a volume integral with the so-called PML-normalization approach [3,4]. Sections 5.1 and 5.2 provide more details, see also [7]. Approximately 20 COMSOL models that can be directly used with **QNMEig** are provided in the *MANMODELS* folder. They cover many geometries and materials (see Table 1) of current interest or can be adapted to cope with the user applications.

### 2.3.2 QNMPole

The QNM solver **QNMPole** has virtually nothing to do with **QNMEig**. It computes and normalizes the QNMs by searching poles in the complex frequency plane [5], starting from an initial guess value and iteratively converging towards the pole. It is much more general than **QNMEig**: it can be used for any material dispersion (Table 1) and can be interfaced with any frequency domain Maxwell solver. It also requires less computational skills than **QNMEig**: it does not require auxiliary fields nor Perfectly Matched Layers (PMLs), although it may be interfaced with solvers implementing PMLs. In return, it does not benefit from the enjoyable professionalism and effectiveness of commercial solvers, such the RF Module eigenmode solver of COMSOL or the eigen solver of CST Studio Suite: the QNM computation is not automated, is performed sequentially and, for each new computation, and initial guess of $\widetilde{\omega}$ has to be set by the user. In fact, **QNMPole** is intentionally maintained at a primitive stage of development for the sake of versatility and generality.

Let us recap the principle of operation:

Step 1. Using their favorite Maxwell solver, the user sets a time-harmonic oscillating dipole source $\mathbf{J}(\omega)$ located in the vicinity of the resonator and computes the field $\mathbf{E}_S(\mathbf{r},\omega)$ scattered by the resonator for three guessed frequencies, $\omega_1$, $\omega_2$ and $\omega_3$, that are slightly different from the searched QNM frequency (pole). These frequencies are complex valued. It may happen that, with some commercial software, e.g., COMSOL, the computation is constrained to real-valued $\omega$'s only. This pitfall is easily circumvented by a simple trick that consists in renormalizing the permittivity and permeability, see the userguide QNMPole V8.pdf or Appendix 1 in [5].

Step 2. The user selects a field component, say $E_{S,z}(\mathbf{r}_t)$, at a fixed position $\mathbf{r}_t$. Step 1 has generated three complex-valued fields, $E_{S,z}(\mathbf{r}_t,\omega_1), E_{S,z}(\mathbf{r}_t,\omega_2), E_{S,z}(\mathbf{r}_t,\omega_3)$. From this initial step, a new estimate $\omega_4$ of the actual pole $\widetilde{\omega}$ is generated (Eq. 20 in Appendix 2 in [5])

$$\omega_4 = \frac{\omega_1(\omega_2-\omega_3)/Z_1 + \omega_2(\omega_3-\omega_1)/Z_2 + \omega_3(\omega_1-\omega_2)/Z_3}{(\omega_2-\omega_3)/Z_1 + (\omega_3-\omega_1)/Z_2 + (\omega_1-\omega_2)/Z_3},$$ (2)

where $Z_p = 1/E_{S,z}(\mathbf{r}_t, \omega_p)$, $p = 1,2,3$.

Step 3. The user computes the field $\mathbf{E}_S(\mathbf{r}, \omega_4)$, and a new value $Z_4 = 1/E_{S,z}(\mathbf{r}_t, \omega_4)$ is obtained. From the quadruplet $(\omega_1, \omega_2, \omega_3, \omega_4)$, the frequency corresponding to the largest $|Z|$ is eliminated and, with the new triplet, one goes back to step 2 and iterates.

Step 4. (Final step) The iterative procedure convergences very rapidly and after typically three iterations (Fig. 2), the new estimated frequency $\omega_n$ is very close to the actual pole $\widetilde{\omega}$ ($|Z_n| \ll 10^{-10}$). The user then computes $\omega_{n+1}$ using step 2 and generate the normalized QNM according to the following formula (Eq. 8 in [5])

$$\widetilde{\mathbf{E}}(\mathbf{r}) = \left(\frac{\omega_{n+1}-\omega_n}{-i\mathbf{J}(\omega_n)\cdot\mathbf{E}_S(\mathbf{r}_0,\omega_n)}\right)^{1/2} \mathbf{E}_S(\mathbf{r},\omega_n). \tag{3}$$

with $\mathbf{r}_0$ the location of the source $\mathbf{J}$.

Clearly, the use of **QNMPole** to compute and normalize QNMs is very general and simple: it just requires knowing an initial guess value for the QNM eigenvalue and solving Maxwell equations a few times for a few frequencies.

In practice, it is convenient to automatize the iterative procedure, implying that the frequency-domain software should be driven by an external program. For most software, the automatization is straightforwardly implemented. In **MAN**, an example of automatization is provided with the RF Module of COMSOL, see MATLAB program Script_QNM_web.m of the *TUTORIALS/Pole* folder.

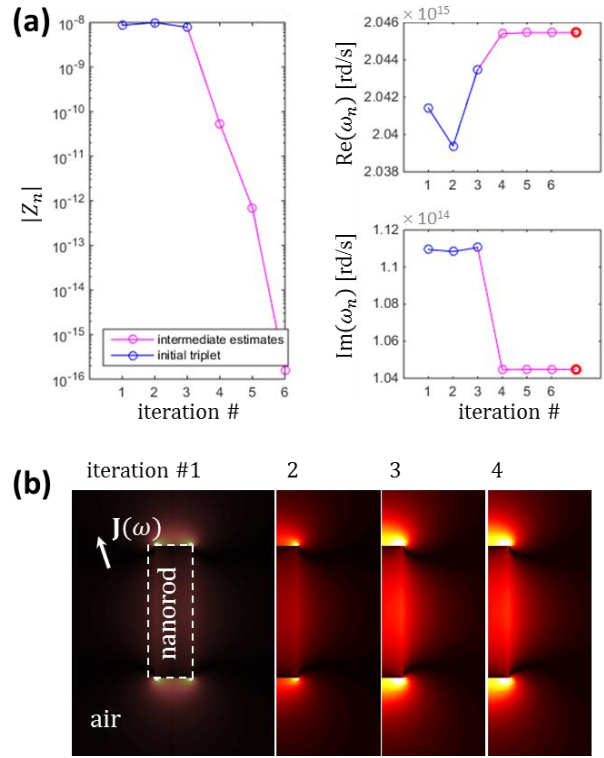

**Fig. 2.** Typical convergence observed with the **QNMPole** for a metallic nanorod. (**a**) The three guessed frequencies are labeled with blue circles. Further iterations are represented with pink circles. Left panel: inverse of the scattered field computed at every iteration at the evaluation point $\mathbf{r}_t$; the inverse value should decrease by several orders of magnitude to guarantee convergence as one approaches the pole $\widetilde{\omega}$. Right panel: Convergence of $\text{Re}(\omega_n)$ and $\text{Im}(\omega_n)$ as one approaches the pole. A plateau is already observed for the first iteration. (**b**) Convergence towards the normalized QNM. More details in [5].

### 2.3.3 QNMPole vs QNMEig: which one should I choose?

**QNMEig** relies on a professional tool that computes a large number (set by the user) of QNMs automatically without any guess values. We recommend the use of **QNMeig** in general.

However, the user has to be experienced, ideally aware of the COMSOL Multiphysics Weak-form Module used to linearize the eigenproblem with auxiliary fields. Difficulties may be also encountered for certain dispersive materials that cannot be simply described with a few poles of a Lorentz-Drude or N-pole Lorentz permittivity.

To help the user, several documents (click Sphere, Cubesubstrate, … in the MAN User Guide.html file in the *DOC* folder) are provided; they explain how these COMSOL models are built. Special attention has been paid to explain how to handle dispersive media by coupling the COMSOL RF and Weak-form Modules. The step to build a QNMEig model is also facilitated by the many models available in the *MANMODELS* folder. These models include optical cavities of different geometries and materials (Si, Au, Ag, AlGaAs, …). The users are encouraged to build their own models by modifying the existing examples that are provided.

**QNMPole** is recommended for users not familiar with COMSOL. The building of the models is fairly easy, as they are very similar to the COMSOL models used for simulating the scattered field at real frequency. It is also very general and can be implemented with any geometries or material dispersions. The counterpart is that **QNMPole** computes and normalizes QNMs one by one and an initial guess value of the eigenfrequency should be provided by the user. Furthermore, the COMSOL models need to be operated by a MATLAB script to implement the iterative scheme of Section 2.3.2. **QNMPole** is recommended for users beginning in computational electromagnetism and when only a few dominant QNMs are required.

### 2.4 QNM expansion

QNM expansion consists in decomposing the resonator response as a sum of QNM contributions. It is central in any resonator analysis and is implemented in several toolboxes of **MAN**, e.g., the *reconstruction* toolbox (see Section 4 for details).

The response of any resonator can be expended (i.e., the field $\mathbf{E}_S$ scattered by the resonator) into a sum of QNM contributions [1], either in the frequency domain $[\mathbf{E}_S(\mathbf{r}, \omega), \mathbf{H}_S(\mathbf{r}, \omega)] = \sum_m \alpha_m(\omega) [\tilde{\mathbf{E}}_m(\mathbf{r}), \tilde{\mathbf{H}}_m(\mathbf{r})]$, or assuming that the excitation pulse can be Fourier transformed, in the time domain $[\mathbf{E}_S(\mathbf{r}, t), \mathbf{H}_S(\mathbf{r}, t)] = \text{Re}\left(\sum_m \beta_m(t) [\tilde{\mathbf{E}}_m(r), \tilde{\mathbf{H}}_m(r)]\right)$. The toolbox for reconstructing the scattered field in the time domain is under construction at the time the publication is written.

The literature distinguishes two classes of methods to expand the monochromatic response (i.e., the field scattered by the resonator excited by a monochromatic driving field at frequency $\omega$) into a sum of QNM contributions, see Table 1 in Ref. [1].

The residue-expansion method exploits the fact that, under certain conditions, a representative function of the system response (often the Green's tensor) is a meromorphic function that can be decomposed as a sum of its poles (Mittag-Leffler theorem), which are the QNMs [19]. Completeness of the QNM expansion of the scattered field, $[\mathbf{E}_S(\mathbf{r}, \omega), \mathbf{H}_S(\mathbf{r}, \omega)] = \sum_m \alpha_m(\omega) [\tilde{\mathbf{E}}_m(\mathbf{r}), \tilde{\mathbf{H}}_m(\mathbf{r})]$, is achieved for a subspace of $\mathbb{R}^3$ (inside the resonator) and for resonators surrounded by a uniform background (no substrate) [3].

In the second method adopted in **MAN**, the continuous Maxwell operator, originally defined on an open space, is replaced by a discretized linear operator (a matrix), with a bounded physical domain bounded by perfectly matched layers (PMLs) [20]. The latter transforms the open space into a regularized Hilbert space [3,7,21], replacing the exponential growth of the QNM field in the open space by an exponential decay in the PMLs. The regularized eigenvectors of the discretized operator are composed of a finite subset of the true QNMs $[\tilde{\mathbf{E}}_m(\mathbf{r}), \tilde{\mathbf{H}}_m(\mathbf{r})]$ of the open system (typically the subset includes the dominant resonances driving the resonator response) and a large, albeit finite, subset of numerical modes (sometimes called PML modes [21]), which originates from the finite-discretized space bounded by PMLs [3,7]. The QNM expansion

$$\begin{cases} [\mathbf{E}_S(\mathbf{r},\omega), \mathbf{H}_S(\mathbf{r},\omega)] = \sum_m \alpha_m(\omega) [\widetilde{\mathbf{E}}_m(\mathbf{r}), \widetilde{\mathbf{H}}_m(\mathbf{r})] + \sum_p \alpha_p^{num}(\omega) [\widetilde{\mathbf{E}}_p^{num}(\mathbf{r}), \widetilde{\mathbf{H}}_p^{num}(\mathbf{r})] \\ \alpha_m \text{ and } \alpha_p^{num} \text{ being known with closed} - \text{form expressions} \end{cases}, \quad (4)$$

is complete for all **r** of the regularized (numerical) space, including the PMLs (see Section 5.3 for detail).

In Eq. (4), the numerical modes $[\widetilde{\mathbf{E}}_p^{num}(\mathbf{r}), \widetilde{\mathbf{H}}_p^{num}(\mathbf{r})]^T$ are denoted just like the true QNMs, by simply adding a superscript '$num$' for the sake of differentiation. They are computed exactly as the true QNMs, by solving the source-free Maxwell equations of Eq. (1). The numerical modes originate from finite-thickness PMLs and replace the continuous spectrum of the operator resulting from the coupling with the open space. Unlike the QNMs, the eigenfrequencies and mode profiles of these modes are sensitive to PML-parameter variations, e.g., the PML damping parameter and thickness. This is the reason why these modes are called "numerical modes".

The dependence of the excitation coefficients, $\alpha_m$ or $\alpha_p^{num}$, with the driving field is analytical, see Sections 4.2.1 and 4.2.2. This implies that, once the eigenvectors are computed and normalized, the reconstruction of the scattered field for many instances of the driving field becomes trivial, even in the temporal domain for resonators illuminated by pulses (Section 4.4).

The true QNMs and the numerical modes are treated in **MAN** in the same way; they are computed with the same solvers, and their excitation coefficients obey the same closed-form expression. In fact, mathematically, it is therefore not necessary to distinguish them, and we will abusively refer to both types of eigenvectors as QNMs, hereafter.

## 3 Getting started

### 3.1 Installation

To install MAN, one needs to start MATLAB and then run the function *addManPath.m* in the main directory:

```
>> addManPath;
```

The function will add the path of the main director as well as the paths of all subdirectories to the MATLAB path.

### 3.2 Quick start

For a quick start, run the Tutorial1D.m file in the *TUTORIALS/Common* folder. Tutorial1D.m provides a step-by-step illustration of how to compute and normalize QNMs with **QNMPole** for a very simple example: a 1D slab (Fabry Perot) in air.

This MATLAB program can be run independently of COMSOL. It illustrates how does **MAN** helps the user to identify the QNMs that are dominantly responsible for the resonant response. This is often an important step in any physical study, because the complex plane contains many QNMs and numerical modes, and not all of them are significant from a physical point of view. In **MAN**, a great attention has been devoted to sort the QNMs by evaluating their excitation probability by a plane wave and distinguishing true QNMs from numerical modes (details in Fig. 3).

Furthermore, the Tutorial1D.m file pedagogically illustrates how to interface **QNMPole** with their own frequency-domain Maxwell solver.

### 3.3 Tutorial for advanced geometries

For studying advanced geometries, one needs to start COMSOL Multiphysics with MATLAB. The *MANMODELS* folder contains more than 20 COMSOL models (.mph files). These models are built with COMSOL 6.1. Users running older versions of COMSOL may have trouble in opening and loading these models. To avoid the issue, they may simply run the buildAndSaveComsolModels.m file in the *TUTORIALS* folder:

```
>> buildAndSaveComsolModels;
```

This function will automatically remove all the existing .mph files of the *MANMODELS* folder and rebuild models compatible with the COMSOL version of the user.

The user may then run the files Tutorial2D.m or Tutorial3D.m in the *TUTORIALS/Common* folder. These files that are much similar to the Tutorial1D.m file and provide a step-by-step illustration of how to compute, normalize, sort, and visualize QNMs using **QNMEig** or **QNMPole**.

The Tutorial2D.m or Tutorial3D.m files are quite similar. We consider Tutorial2D.m for a rapid computation:

1. Select **QNMEig** or **QNMPole** solver to compute QNMs.

```
>>mansolver = 'eig'; or >>mansolver = 'pole';
```

2. Select a model in the MANMODELS folder. Models begining by 'QNMEig' and 'QNMPole' have to be used with **QNMEig** (mansolver = 'eig') and **QNMPole** (mansolver = 'pole'), respectively.

```
>>name = 'QNMEig_2DSiWireTM.mph';
```

3. One may plot the geometry under study by calling

```
>>plotGeometry2D(Model,'display_units','micro');
```

4. The MATLAB script then loads the selected .mph file.

```
>>Model = loadModels(name,'solver', mansolver);
```

5. The MATLAB script either drives COMSOL to compute the QNMs or loads the precomputed results if the model has already been used in a previous study

```
>>QNM = getEigQNMs(Model);
```

The function getEigQNMs first checks whether the .mph file contains precomputed QNMs in its dataset. If not, it will drive the COMSOL to compute the QNMs and then load them.

6. In general, the QNM computation will consume most of the computer time (especially in 3D). The user may save the results in the data set of the COMSOL file:

```
>> [mphfile,mphfolder] = saveModel(Model);
```

The QNMs will be loaded if the model is used gain another time.

7. Once the QNMs are computed or loaded, the next step is to determine the QNMs of physical interest. Important information about the QNMs is displayed using the function plotComplexPlane

```
>> plotComplexPlane(Model,QNM);
```

Figure 3 displays the MATLAB figure generated by the function; the figure displays the complex eigenfrequencies of all the computed eigenstates in the complex frequency plane. In general, the user wishes to first identify the dominant QNMs. It is often a tedious task, and **MAN** helps the user by introducing two figures of merit:
- the mode ratio ($MR \in [0,1]$, see the MAN User Guide.html in the *DOC* folder for the precise definition) shown by the color of the disk; $MR$ indicates whether the eigenstate is a true QNM ($MR = 1$, yellow) or a numerical mode ($MR = 0$, blue);
- the excitation strength ($ES \in [0,1]$, see the MAN User Guide.html for the precise definition) shown by the size of the disc; $ES$ indicates the modal excitation strength for an incident plane wave (large discs correspond to large excitation strengths).

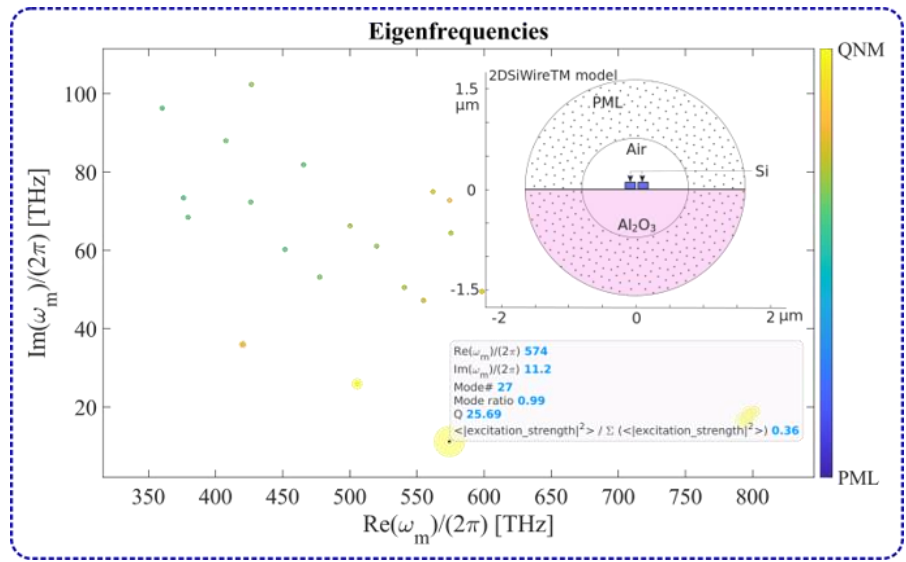

**Fig. 3**. MATLAB figure provided by the plotComplexPlane.m function for a coupled pair of 2D silicon nanowires (QNMEig_2DSiWireTM.mph). It shows the complex frequency plane and displays important information on the QNMs: the size and color of the markers visualize the excitation strength and mode ratio, respectively. Other important information about the QNMs can be shown by clicking the discs. Inset: Sketch of the nanowire pair.

8. Finally, the user may load the QNM near fields and permittivities in any user-defined coordinates by calling the functions,

```
>> QNM = getModalFields(Model, QNM,'coord',coord);
>> permitt = getRelativePermittivity(Model, QNM,coord);
```

to further plot maps of the QNM fields and the permittivity of the geometry.

### 3.4 What to do if the user does not have a COMSOL license

Most QNM solvers and toolboxes in **MAN** are interfaced with COMSOL. This is inconvenient for those who use other Maxwell solvers. Therefore, **MAN** includes a tutorial example, which can be run with MATLAB only and illustrates how to interface the solver **QNMPole** with a frequency-domain Maxwell solver. The tutorial example, Tutorial1D.m (Section 3.2), can be found in the *TUTORIALS/Common* folder.

Tutorial1D.m is easy to read. It relies on a trivial Maxwell solver that computes the field scattered by a 1D slab excited by a planar source. Quite simply, the 1D Maxwell solver can be replaced by other frequency-domain 2D or 3D Maxwell solvers. The user only needs to modify the following two interface functions:

- Mfun.m: this MATLAB function drives a frequency-domain Maxwell solver to compute the scattered fields at complex frequency. It returns the scattered fields at the test point $\mathbf{r}_t$ and the driving source location $\mathbf{r}_0$ (see Section 2.3.2), and a file (the file type depends on the Maxwell solver) used by the second interface function, fMawell_toy_post.m. In the present 1D example, the file is a .mat file that contains the amplitudes of the plane waves that determine the scattered fields inside and outside the slab.
- fMaxwell_toy_post.m: this MATLAB function performs a post-processing treatment of the file generated by Mfun. It returns the scattered fields and permittivity in any user-defined coordinates.

Note that not all commercial Maxwell solvers, e.g. the solver of the COMSOL RF Module, do not handle complex frequencies. The user may refer to Appendix 1 in [5] where a simple trick is provided to circumvent this difficulty.

# 4 Toolboxes

## 4.1 Overview

Table 2 lists the toolboxes of Version 8. New toolboxes may be added in future versions for other applications. This section provides a short introduction to the toolboxes content; a step-by-step presentation can be found in the user guide documents inside the *DOC/pdf/toolboxes* folder.

**Toolbox organization.** Each toolbox can be found in a separate folder, which is composed of a user guide document and several dedicated MATLAB source codes. Most toolboxes rely on general MATLAB functions found in the *SOLVER* folder to generate figures, load QNM fields and other information from the COMSOL models; thus the *SOLVER* folder should be added to the MATLAB path before using the toolboxes.

The toolboxes use the COMSOL models of the *MANMODELS* folder to compute the QNMs. To study other optical systems, new COMSOL models should be implemented by the user. Since the toolboxes often rely on parameters defined in the COMSOL models, e.g., the parameters of the Drude-Lorentz functions, the background permittivity, and the domain numbers defining the resonator, the user is recommended to follow the instruction of MAN User Guide.html and get inspired from the available models to build their own models to accurate access to these parameters.

**Caution.** The toolboxes have been developed over different periods of time; their level of development may differ. For example, in the *reconstruction* toolbox, the user only needs to provide the angle of the incident plane wave and the background layered stack; the background field inside the resonator (required to further compute the QNM excitation coefficients) is automatically computed. However, in the *nonlinear* toolbox, an analytical formula has to be provided for the background field inside the resonator. When starting with **MAN**, it is recommended to first study the code examples that are provided before developing a new application.

**Table 2.** List of the toolboxes in MAN V8. (*) denotes the toolboxes released in V8 for the first time.

| Toolbox acronym (Year of first release) | Further reading | Description |
|---|---|---|
| *reconstruction* (*) (2022) | [4,5,7] | Reconstruct the electromagnetic fields, absorption and extinction cross-sections, Purcell factors, and LDOS using QNMs in the frequency domain. Compute the QNM excitation coefficients $\alpha(\omega)$. |
| *alpha-coefficient* (2019) | [7] | Same as for the reconstruction toolbox, except that only resonators embedded in free space and excited with a plane wave can be analyzed. |
| *nonreciprocal* (2021) | [12] | Compute and normalize QNMs for resonators made of nonreciprocal materials. |
| *nonlinear* (2020) | [10] | Show how the second-harmonic generation can be analyzed in the QNM basis. |
| *grating* (2019) | [8] | Compute the specular reflection of 1D gratings using QNM expansions. |
| *far-field-radiation* (*) (2022) | [22] | Generate the far-field radiation diagrams of QNMs. |
| *multipole* (2020) | [9] | Compute the intrinsic multipole moments of QNMs for resonators in free space. |
| *interpolation* (2021) | [11] | Provide a fast and accurate reconstruction of the resonator response by interpolating the non-resonant contribution at real frequency. |
| *axisymmetric* (*) (in preparation) | | Same as the reconstruction toolbox for optical resonators with axial symmetry. |

| *Temporal domain* (in preparation) | [7,23] | Reconstruction in the temporal domain, for instance for a pulse excitation (Section 4.4). |

**4.2 Near-field reconstruction in the frequency domain**

In the *reconstruction* and the *alpha-coefficient* toolboxes, the reconstruction is performed with the QNM expansion of Eq. (4) over the regularized space, including the PMLs. If all the eigenvectors are considered in the expansion, the reconstruction is exact, in the sense that the reconstructed field obtained with the sum of all QNM contributions is mathematically equal to the field computed with a real frequency solver with the same mesh [7].

An important question is how fast one approaches the exact result by increasing the number $N$ of QNMs retained in the expansion. We may be interested in two kinds of convergence speed: asymptotic behaviors as $N \to \infty$ or initial behaviors on the accuracy achieved with only a few dominant QNMs (typically $N < 10$), which govern the physics of the system. The first kind of convergence is more mathematically directed, and the second kind is very important in practice. So far, the whole literature provides no clear answer to the questions of the convergence issues. Only a few numerical studies have been performed from case to case. State-of-the-art studies of the convergence performance for landmark plasmonic resonators can be found in Ref. [7]. Other interesting cases can be found in Refs. [8,21,24-26].

The convergence issue is exacerbated by the existence of an infinite number of rigorous formulas for the $\alpha_m$'s (or $\alpha_p^{num}$'s) for dispersive materials. The non-uniqueness arises from an infinite choice for implementing the source term with auxiliary fields, see Ref. [26] for details. Close to the resonance frequency ($\widetilde{\omega}_m \to \omega$), all the formulas converge to the same value, see Section 4.2 in [1]. However, away from the resonance, the formulas give rise to significantly different values of $\alpha_m$'s. The authors believe that the choice of the source term has a weak impact on the asymptotic convergence rate. However, it has an impact on the accuracy achieved for small $N's$.

Over the past five years, some of the authors have studied this accuracy for various nanoresonator geometries. Depending on the geometry, they observe either very similar [26] or markedly different reconstruction accuracies [9,10]. They could not establish general rules, and this is the reason why special attention has been devoted to the implementation of the *reconstruction* toolbox. Overall, the latter implements a quite complete overview of the main methods available for reconstructing the near fields. The user may first follow the approach in [7] (method **M 1** hereafter). Then, if the reconstruction accuracy is not satisfactory, other methods, especially method **M 2** that was successfully used in recent works [9,10], can be tried.

The *reconstruction* and *alpha-coefficient* toolboxes also compute useful optical responses, e.g., the extinction and absorption cross-sections, and the Purcell factors. Note that, in contrast with the *reconstruction* toolbox, the *alpha-coefficient* toolbox is valid only for resonators in free space and does not reconstruct the near-field. It can be viewed as a simplified version of the *reconstruction* toolbox for resonators in uniform media (without substrate).

**4.2.1 QNM excitation coefficient**

For non-dispersive materials, there is a unique expression for the $\alpha_m$'s (or $\alpha_p^{num}$'s) [5,26]

$$\alpha_m(\omega) = \iiint_{V_{res}} \left( \Delta\varepsilon(\mathbf{r},\omega) \frac{\omega}{\widetilde{\omega}_m - \omega} \right) \mathbf{E}_b \cdot \widetilde{\mathbf{E}}_m d^3\mathbf{r}, \tag{5}$$

where $\Delta\varepsilon(\mathbf{r},\omega) = \varepsilon(\mathbf{r},\omega) - \varepsilon_b$ is the permittivity difference defining the resonator volume $V_{res}$ in the scattering field formulation, see Annex 2 in Ref. [1] for details, and $\mathbf{E}_b$ is the background field (the driving field computed for $\Delta\varepsilon = 0$). Note that a $\omega$-dependence is included in $\Delta\varepsilon(\mathbf{r},\omega)$ in Eq. (5) to account for the fact that the expression is also valid for dispersive materials.

For resonators with Drude, multi-pole Drude-Lorentz, or multi-pole Lorentz permittivities, a classical and well-documented expression is [7]

$$\alpha_m(\omega) = \iiint_{V_{res}} \left( \varepsilon_b(\mathbf{r},\omega) - \varepsilon_\infty(\mathbf{r}) + \Delta\varepsilon(\mathbf{r},\widetilde{\omega}_m) \frac{\widetilde{\omega}_m}{\widetilde{\omega}_m - \omega} \right) \mathbf{E}_b \cdot \tilde{\mathbf{E}}_m d^3\mathbf{r}, \tag{6}$$

where $\varepsilon_b(\mathbf{r},\omega)$ is the background permittivity and $\varepsilon_\infty(\mathbf{r})$ is the permittivity $\varepsilon(\mathbf{r},\omega \to \infty)$. Note that for non-dispersive materials, Eq. (5) and Eq. (6) are identical. $\varepsilon_b(\mathbf{r},\omega)$ does not necessarily correspond to a homogeneous medium. It may correspond to a layered substrate (Fig. 4). In this case, the background field $\mathbf{E}_b$ is composed of counterpropagating plane waves that are reflected and transmitted in the layers. It can be computed with 2 × 2 matrix products for isotropic materials; this computation is automatically performed by the *reconstruction* toolbox thank to the freeware RETICOLOfilm-stack [27] provided that the user defines the background permittivities and the incident plane-wave parameters.

According to our experience, the formulas of Eqs. (3) and (4) provide similar convergence performance when reconstructing the total field using the methods in Table 3.

Expressions for $\alpha_m$'s also exist for resonators driven by localized sources, such as electric (**p**) or magnetic (**m**) dipoles located at a position $\mathbf{r}_0$ in the near field. An infinity of expressions may also be derived for dispersive resonators; however, in contrast to the plane wave case, the whole literature seems to have adopted the convenient (fully analytical) expression [4,12]

$$\alpha_m^{(t)}(\omega) = \frac{\omega}{\widetilde{\omega}_m - \omega} [\mathbf{p} \cdot \tilde{\mathbf{E}}_m(\mathbf{r}_0, \widetilde{\omega}_m) - \mathbf{m} \cdot \tilde{\mathbf{B}}_m(\mathbf{r}_0, \widetilde{\omega}_m)], \tag{7}$$

for the total field $[\mathbf{E}_{tot}(\mathbf{r},\omega), \mathbf{H}_{tot}(\mathbf{r},\omega)] = \sum_m \alpha_m^{(t)}(\omega)[\tilde{\mathbf{E}}_m(\mathbf{r}), \tilde{\mathbf{H}}_m(\mathbf{r})]$, see Section 5.2 in [1]. The *reconstruction* toolbox implements this expression only.

### 4.2.2 Field reconstruction

To illustrate the possibilities offered by the *reconstruction* toolbox, consider the representative system shown in Fig. 4. The system is composed of five domains corresponding to five different materials. The background is composed of three domains, labeled 2, 3, and 4. The resonator is composed of two other domains. We assume that only the materials of domains 1 and 4 are dispersive.

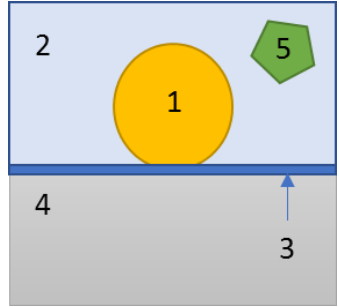

**Fig. 4.** Illustrative example of a system suited for simulation with the *reconstruction* toolbox. The background is defined by a multilayer structure (domains 2, 3, 4). The resonator is defined by the two other domains, 1 and 5. Only the materials of domains 1 and 4 are dispersive.

The QNMs have different field components, and each component can be used for the reconstruction of the total electric field $\mathbf{E}_{tot}$ (see Section 5.4). The *reconstruction* toolbox implements 2×4 different methods (labeled **M 1**, **M 2**, **M 3**, and **M 4** in Table 3) to reconstruct $\mathbf{E}_{tot}$ for illumination by a plane wave. The factor 2 accounts for the fact that two different equations for the $\alpha_m$'s can be used for each method. The methods have been selected for their accuracy, simplicity, or broad use.

For all methods, the reconstruction in the non-dispersive domains (2, 3, 5 in the example) is always performed with

$$[\mathbf{E}_{tot}(\mathbf{r},\omega), \mathbf{H}_{tot}(\mathbf{r},\omega)] = [\mathbf{E}_b(\mathbf{r},\omega), \mathbf{H}_b(\mathbf{r},\omega)] + \sum_m \alpha_m(\omega)[\tilde{\mathbf{E}}_m(\mathbf{r}), \tilde{\mathbf{H}}_m(\mathbf{r})], \tag{8}$$

$\alpha_m(\omega)$ being given by Eq. (5) or Eq. (6). The methods differ only for the dispersive domains, 1 and 4.

**Table 3.** A summary of important formulas implemented in the *reconstruction* toolbox for reconstructing the total field, $\mathbf{E}_{tot}$, in dispersive domains. Four methods (**M 1**-**M 4**) are considered, and each may rely on two different formulas for the $\boldsymbol{\alpha_m}$'s.

| $\alpha_m$ | Eq. (6) | | Eq. (5) | |
|---|---|---|---|---|
| Dispersive Dom. n° | 1 (resonator) | 4 (background) | 1 (resonator) | 4 (background) |
| **M 1** | $\sum \alpha_m(\omega) \tilde{\mathbf{E}}_m + \mathbf{E}_b$ | | | |
| **M 2** | $\sum \alpha_m(\omega) \frac{\varepsilon(\tilde{\omega}_m) - \varepsilon_\infty}{\varepsilon(\omega) - \varepsilon_\infty} \tilde{\mathbf{E}}_m$ | | $\sum \alpha_m(\omega) \frac{\varepsilon(\tilde{\omega}_m) - \varepsilon_\infty}{\varepsilon(\omega) - \varepsilon_\infty} \tilde{\mathbf{E}}_m + \mathbf{E}_b$ | |
| **M 3** | $\sum \alpha_m(\omega) \frac{\varepsilon(\tilde{\omega}_m) - \varepsilon_\infty}{\varepsilon(\omega) - \varepsilon_\infty} \frac{\tilde{\omega}_m}{\omega} \tilde{\mathbf{E}}_m$ | | $\sum \alpha_m(\omega) \frac{\varepsilon(\tilde{\omega}_m) - \varepsilon_\infty}{\varepsilon(\omega) - \varepsilon_\infty} \frac{\tilde{\omega}_m}{\omega} \tilde{\mathbf{E}}_m + \mathbf{E}_b$ | |
| **M 4** | $\sum \alpha_m(\omega) \frac{\varepsilon(\tilde{\omega}_m)}{\varepsilon(\omega)} \frac{\tilde{\omega}_m}{\omega} \tilde{\mathbf{E}}_m + \frac{\varepsilon_b(\omega)}{\varepsilon(\omega)} \mathbf{E}_b$ | | | |

All the methods asymptotically converge towards the same value if all the eigenvectors are included in the expansion. However, for small truncation ranks, they may provide significantly different accuracies. Figure 5 illustrates our purpose for a simple plasmonic system, a Dolmen composed of three Drude silver nanorods embedded in an air background. The response of the Dolmen at optical frequencies is dominated by four QNMs shown in Fig. 5(b). Figure 5(c) shows the reconstruction performed with methods **M 1-4** and $\alpha_m(\omega)$ computed with Eq. (6) for a wavelength of 700 nm. They are obtained for an incident field polarized along the $x$-direction, $\mathbf{E}_b = -\mathbf{e}_x \exp(ik_b z)$, and are compared with Fig. 5(d) that displays the full-wave frequency simulated result obtained with COMSOL. For the reconstruction inside the Dolmen, methods **M 2**-**M 4** are more accurate compared to the most intuitive method **M 1**. The authors have also observed a faster convergence of **M 2**-**M 4** for other nanostructures, such as a Si nanocylinder on a low refractive index substrate [10]. They refer the readers to Section 5.4 for a possible explanation of the faster convergence of **M 2**-**M 4** for a small set of QNMs.

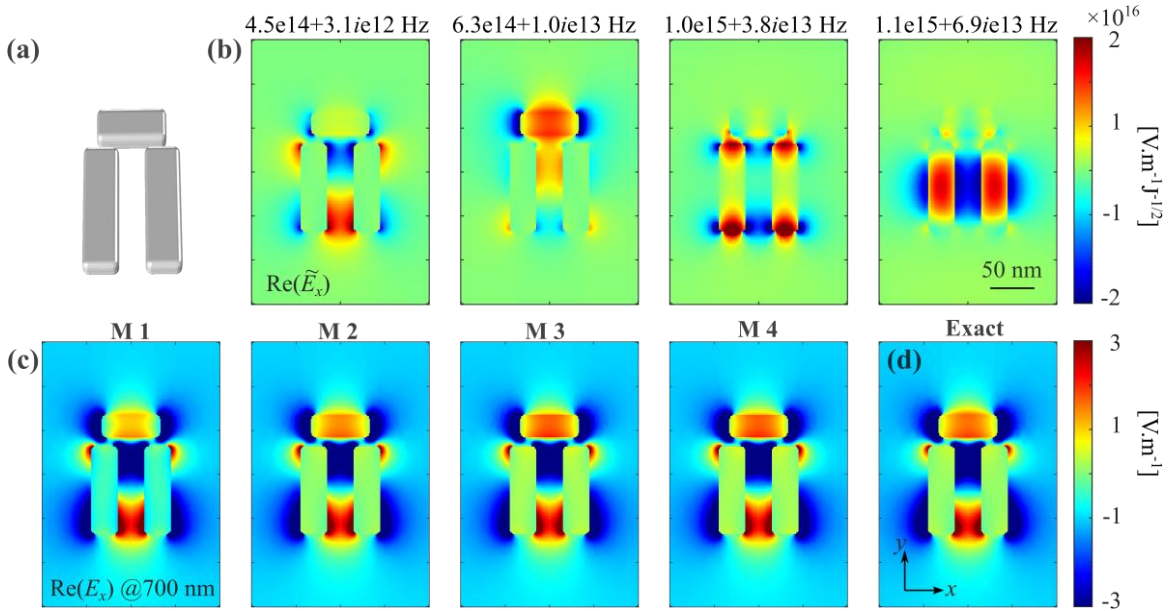

**Fig 5.** QNM reconstruction of the real part of the electric field, $\text{Re}(E_x)$, for a Dolmen-like structure. (**a**) The plasmonic Dolmen is made of silver and composed of an upper rod ($66 \times 26 \times 20$ nm$^3$) separated by a gap of width $g = 10$nm from two lower rods ($30 \times 100 \times 20$ nm$^3$) separated by 30 nm. The silver permittivity of the nanorods is approximated by a single-pole Drude model with

$\varepsilon_\infty = 1$, $\omega_p = 1.366 \times 10^{16}$ rad, and $\gamma = 0.0023\omega_p$. The Dolmen is embedded in air ($\varepsilon_b = 1$). **(b)** Electric field $x$-component $\text{Re}(\tilde{E}_x)$ in $xy$-plane of the four dominant QNMs used in the reconstruction. **(c)** Real part of the scattered electric fields $\text{Re}(E_x)$ for $\lambda = 700$ nm reconstructed with methods **M 1**-**M 4**. The incident field is a plane wave polarized along the $x$-direction. **(d)** $\text{Re}(E_x)$ obtained with fully vectorial numerical computations. The user is encouraged to repeat the figure by running QNMToolbox_reconstruction_Dolmen2.m in the *TOOLBOXES/ QNMtoolbox_reconstruction* folder.

### 4.2.3 Extinction and scattering cross-sections, Purcell factor

In the *reconstruction* and *alpha-coefficient* toolboxes, the extinction and scattering cross-sections, $\sigma_{Ext}$ and $\sigma_{Abs}$, are accurately computed with the following formulas [5]

$$\sigma_{Ext} = -\frac{\omega}{2I_0} \iiint_{V_{res}} \text{Im}[\Delta\varepsilon(\omega)\mathbf{E}_{tot} \cdot \mathbf{E}_b^*]dV, \tag{9}$$

and

$$\sigma_{Abs} = -\frac{\omega}{2I_0} \iiint_{V_{res}} \text{Im}[\varepsilon(\omega)]|\mathbf{E}_{tot}|^2 dV, \tag{10}$$

with $I_0$ the time averaging Poynting vector of the incident plane wave. The scattering cross section is then deduced with the difference $\sigma_{Sca} = \sigma_{Ext} - \sigma_{Abs}$.

Since multiple methods can be used for the reconstruction of $\mathbf{E}_{tot}$, an equal number of expressions for $\sigma_{Ext}$ and $\sigma_{Abs}$ that give different results for truncated expansions are available. The *reconstruction* toolbox implements time-tested expressions [5,7].

For dispersive resonators and provided that the $\alpha_m$'s are computed with Eq. (6), we use the formulas derived in Ref. [7]. The $\Delta\varepsilon(\omega)\mathbf{E}_{tot}$ term in $\sigma_{Ext}$, Eq. (9), is decomposed into two terms: $\Delta\varepsilon(\mathbf{r},\omega)\mathbf{E}_{tot} = (\varepsilon_\infty - \varepsilon_b)\mathbf{E}_{tot} + (\varepsilon(\omega) - \varepsilon_\infty)\mathbf{E}_{tot}$. We use **M 1** (resp. **M 3**) in Table 3 to reconstruct $\mathbf{E}_{tot}$ in the first (resp. second) term. For $\sigma_{Abs}$, we use **M 3** to reconstruct $\mathbf{E}_{tot}$. These formulas may benefit from the fact that **M 3** may offer a faster convergence, as compared to **M 1**, for small sets of QNMs. For non-dispersive resonators, the $\alpha_m$'s are obtained with Eq. (5), $\sigma_{Ext}$ and $\sigma_{Abs}$ are obtained by reconstructing $\mathbf{E}_{tot}$ with method **M 1** [5].

Table 4 summarizes important formulas implemented in the *reconstruction* toolbox to reconstruct the extinction and absorption cross-sections.

**Table 4.** Formulas implemented in the *reconstruction* toolbox to reconstruct $\sigma_{Ext}$ and $\sigma_{Abs}$.

| | $\alpha_m$'s computed with Eq. (5) | $\alpha_m$'s computed with Eq. (6) |
|---|---|---|
| $\sigma_{Ext}$ | $-\frac{\omega}{2I_0} \iiint_{V_{res}} \text{Im}\left[\Delta\varepsilon(\omega)\left(\sum \alpha_m \tilde{\mathbf{E}}_m + \mathbf{E}_b\right) \cdot \mathbf{E}_b^*\right] dV$ | $\iiint_{V_{res}} -\text{Im}\left[\sum [\varepsilon(\tilde{\omega}_m) - \varepsilon_\infty]\tilde{\omega}_m + [\varepsilon_\infty - \varepsilon_b]\omega]\alpha_m \frac{\tilde{\mathbf{E}}_m \cdot \mathbf{E}_b^*}{2I_0}\right] dV$ |
| $\sigma_{Abs}$ | $-\frac{\omega}{2I_0} \iiint_{V_{res}} \text{Im}[\varepsilon(\omega)]\left\|\sum \alpha_m \tilde{\mathbf{E}}_m + \mathbf{E}_b\right\|^2 dV$ | $-\frac{\omega}{2I_0} \iiint_{V_{res}} \text{Im}[\varepsilon(\omega)]\left\|\sum \alpha_m \frac{\varepsilon(\tilde{\omega}_m) - \varepsilon_\infty}{\varepsilon(\omega) - \varepsilon_\infty}\frac{\tilde{\omega}_m}{\omega}\tilde{\mathbf{E}}_m\right\|^2 dV$ |

Figures 6(a-b) show examples of results obtained with the *reconstruction* toolbox for a silver nanocube antenna on a gold substrate, separated by an 8 nm thick polymer spacer. The geometrical and material parameters are found in Fig 6(a). The antenna is illuminated with a plane wave for an incidence angle $\theta_i = 55°$. We adopt Eq. (6) to compute the $\alpha_m$'s. As the antenna materials are dispersive, the *reconstruction* toolbox automatically selects the formulas of the second column of Table 4 to compute $\sigma_{Ext}$ and $\sigma_{Abs}$. The blue (extinction) and red (absorption) curves are obtained by considering two QNMs. For the sake of comparison, we also show reference data (circles) obtained with the frequency-domain solver of COMSOL. The agreement is quantitative over the entire spectrum.

The *reconstruction* toolbox also incorporates functions that allow to compute the Purcell factor $P_m$ and normalized LDOS ($\gamma/\gamma_0$), when a resonator is driven by electric **p** or magnetic **m** dipoles (or a Dirac source with both moments) [4,12]

$$\frac{\gamma}{\gamma_0} = \Sigma_m P_m = -\frac{2}{\hbar\gamma_0}\Sigma_m \text{Im}\left[\alpha_m^{(t)}\left(\mathbf{p}^* \cdot \tilde{\mathbf{E}}_m(\mathbf{r}_0) + \mathbf{m}^* \cdot \tilde{\mathbf{B}}_m(\mathbf{r}_0)\right)\right], \tag{11}$$

where the excitation coefficient $\alpha_m^{(t)}$ is computed using Eq. (7) and $\gamma_0 = \omega^3 [|\mathbf{p}|^2 + |\mathbf{m}|^2/c^2]/(3\pi\varepsilon_0\hbar c^3)$ denotes the vacuum spontaneous decay rate.

Figures 6(c-d) illustrate a simple problem, a gold nanorod driven by an electric dipole emitter oriented parallel to the nanorod axis, for which the Purcell factor is accurately computed with a single QNM in the expansion [4]. Using the classical frequency-domain solver of COMSOL, the normalized LDOS is first computed according to the definition: $\frac{\gamma}{\gamma_0} = -\frac{2}{\hbar\gamma_0}\text{Im}[\mathbf{p}^* \cdot \mathbf{E}_{tot}(\mathbf{r}_0)]$. We then reconstruct the LDOS with Eq. (11). The result reconstructed by considering only the longitudinal electric-dipole QNM is shown with the red line.

Figure 6 provides two examples for which the optical response can be accurately reconstructed with a very few numbers of QNMs. In practice, there also exist systems, e.g. resonators on substrates supporting guiding modes, dipolar emitters placed very close to metal surfaces and quenched, and low Q resonators, for which a large set of QNMs have to be incorporated to achieve an accurate reconstruction [7]. However, we do not recommend the user to compute a large number of QNMs, as it is time consuming. Rather we recommend computing the few QNMs that capture the main resonant feature of the spectrum, and further use the *interpolation* toolbox (Section 4.3) to obtain an accurate reconstruction with small computational loads.

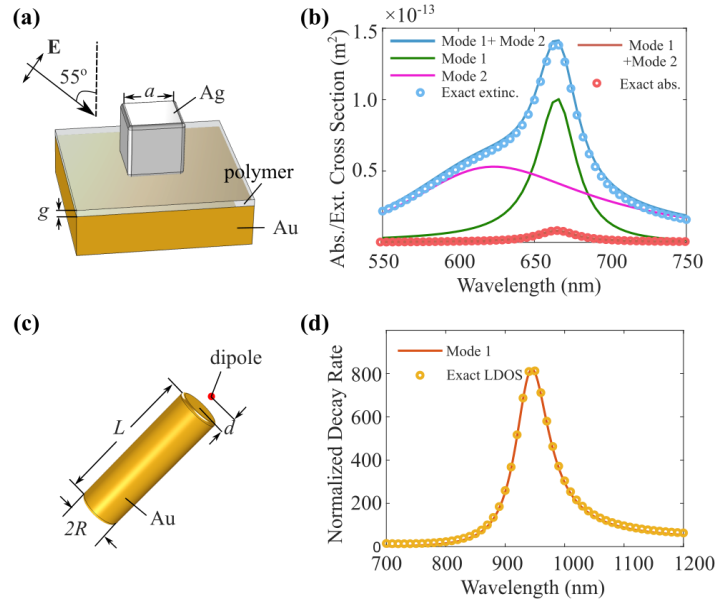

**Fig. 6.** Some well-converged examples of reconstruction in the QNM basis for very small $N$'s. (**a**) Schematic representation of an Ag nanocube antenna of size $a = 65$ nm deposited on a 8-nm-thick polymer film (refractive index $n = 1.5$) coated on an Au substrate. The incident wave is an oblique TM-polarized plane wave. (**b**) Cross-section spectra (solid blue and red curves) reconstructed with 2 QNMs and compared with reference 'exact' data (circles) computed with COMSOL. (**c**) An Au nanorod ($L = 65$ nm, $R = 15$ nm) is embedded in a host medium of refractive index $n = 1.5$. (**d**) Normalized decay-rate spectrum for an on-axis dipole oriented parallel to the nanorod (red dot) and located at a distance $d = 10$ nm. The Ag permittivity is found in the caption of Fig. 5. For gold, we use a double pole Drude–Lorentz model with $\varepsilon_\infty = 6$, $\omega_{p1} = 5.37 \times 10^{15}$ rad, $\gamma_1 = 6.22 \times 10^{13}$ rad, $\omega_{01} = 0$ rad, $\omega_{p2} = 2.26 \times 10^{15}$ rad, $\omega_{02} = 4.57 \times 10^{15}$ rad, and $\gamma_2 = 1.33 \times 10^{15}$ rad. The user is encouraged to repeat the figures by running

QNMToolbox_reconstruction_NPoM_cube.m and QNMToolbox_reconstruction_rod_dipole.m in the *TOOLBOXES/QNMtoolbox_reconstruction* folder.

### 4.3 Effective reconstruction with interpolation

Accurate reconstructions often require considering a large set of QNMs in the expansion in addition to the dominant QNMs of physical significance. For instance, it is systematically the case when the background is not uniform, since branch cuts, modeled as many individual poles, may then significantly contribute to the reconstruction [3,7]. The computation of all these poles (often referred to as numerical modes [3] or 'PML modes' [21]) is usually time-consuming and significantly dilutes the force of the modal approach.

The *interpolation* toolbox implements a simple approach to get around the necessity to consider the numerical modes. It relies on the ansatz that the numerical modes gently contribute to the reconstruction in the spectral range of interest: either their quality factors are extremely small ($< 1$) or their frequencies lie outside the spectral window of interest. The *interpolation* toolbox illustrates how to efficiently interpolate the smooth contribution with only a few ($< 10$) real frequency simulations, to reconstruct the scattered field as a combination of two contributions, that of the few dominant QNMs and that interpolated.

In this way, the physics of the resonator is made transparent and a fast computation with high accuracy is achieved; in general, the computation is much faster than those required with purely real frequency or temporal analysis [11].

### 4.4 Reconstruction in the temporal domain

The study of the dynamics of optical nanoresonators under irradiance by a pulsed beam is indispensable for analyzing ultrafast optical phenomena and nonlinear light-matter interactions [28]. To obtain insights into the physics of the temporal responses, the analysis is best performed in the QNM basis, see Fig. 7.

The *temporal-domain* toolbox takes advantage of the $\omega-$analyticity of the frequency-domain expansion of Eq. (4) to effectively compute the temporal responses of optical nanoresonators in the QNM basis by applying a temporal ($\omega \leftrightarrow t$) Fourier transformation (FT)

$$[\mathbf{E}_S(\mathbf{r},t), \mathbf{H}_S(\mathbf{r},t)] = \text{Re}\left(\sum_m \beta_m(t)[\tilde{\mathbf{E}}_m(\mathbf{r}), \tilde{\mathbf{H}}_m(\mathbf{r})]\right), \tag{12}$$

with $\beta_m(t)$ the time-dependent modal excitation coefficient given by

$$\beta_m(t) = \text{FT}[\alpha_m(\omega)]. \tag{13}$$

In the toolbox, by default, the temporal intensity profile of the incident wave has a Gaussian shape, and the spatial profile is a plane wave. The parameters of the pulsed light, including central frequency, pulse duration, propagation direction and polarization, are set by the user. Two different $\beta_m$'s are computed from Eqs. (3) [6] and (4) [7].

For resonators that can be modeled with a few QNMs, the temporal QNM reconstruction can be extremely advantageous in terms of computational loads. Indeed, once the QNMs are known, the incident pulse duration, polarization or incidence angle can be changed at will since the reconstruction just relies on a 1D Fourier transform, which can be evaluated in a few seconds. In contrast, with the FDTD, any new instance of the driving field requires a new computation. Additionally, the knowledge of the individual contribution of each QNM in the temporal responses significantly helps in interpreting the physics. These two decisive advantages are illustrated in Fig. 7 for a Dolmen-type nanoresonator composed of three gold nanorods. The literature on temporal QNM reconstruction in nanophotonics is scarce; an advanced example of reconstruction with an analysis of the kind of shape that can be obtained for $\beta_m(t)$ can be found in [7].

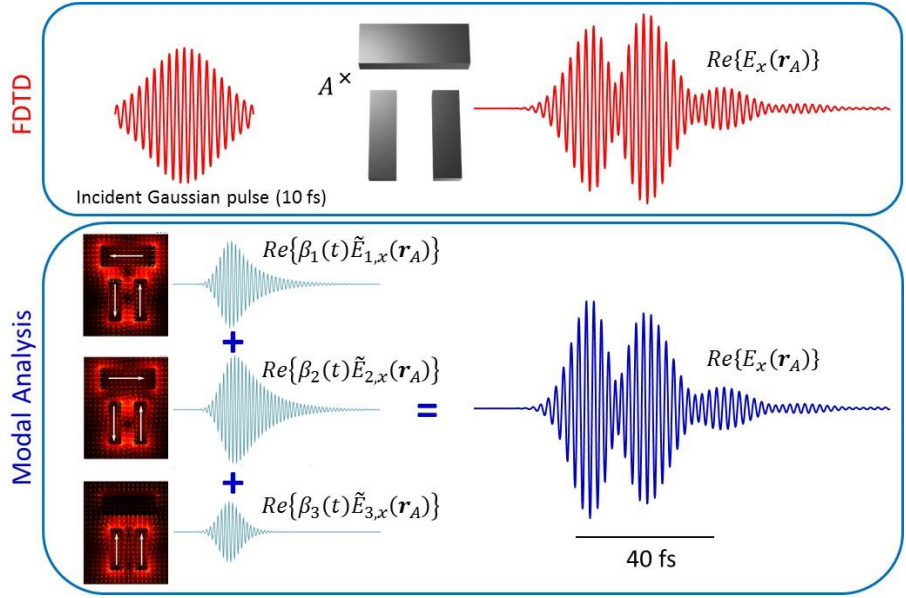

**Fig. 7.** Analysis of the temporal response of nanoresonators with QNM expansions. A Dolmen-type nanoresonator is illuminated by a 10 fs plane-wave Gaussian pulse with a central frequency $\omega_0 = 2.9 \times 10^{15}$ rad/s. Its near-field response at point $A$, as predicted with the FDTD (red), is faithfully reconstructed and explained as resulting from the beating of three dominant QNMs. A comparison between the FDTD data and the 3-QNM reconstruction is provided in Ref. [6].

**4.5 QNM far-field radiation**

The QNMEig and QNMPole solvers compute the near-field pattern of normalized QNMs. For many applications, it is also interesting to know the QNM contribution to the far-field radiation.

The far-field is easily computed by applying a near-to-far-field transformation (NFFT), which computes radiation diagrams. We may symbolically write

$$\text{NFFT}_\omega[\mathbf{E}_S(\mathbf{r}, \omega), \mathbf{H}_S(\mathbf{r}, \omega)] = \sum_m \alpha_m(\omega)\, \text{NFFT}_\omega[\tilde{\mathbf{E}}_m(\mathbf{r}), \tilde{\mathbf{H}}_m(\mathbf{r})]. \tag{14}$$

As shown in Fig. 4d in [7], the reconstruction of the field scattered in the far-field with Eq. (14) is very accurate. Theoretically, the near-to-far-field transform depends on the frequency of interest $\omega$, and this is the reason why we add a subscript $\omega$ to NFFT in Eq. (14). This implies that one should perform a new near-to-far-field transform to know the far-field radiation at each new frequency, say $\omega' \neq \omega$. This may be demanding in computational resources.

To reduce the number of computations, the toolbox performs the near-to-far-field transform for each QNM at its own frequency $\text{Re}(\tilde{\omega}_m)$; the far field then reads as

$$\text{NFFT}_\omega[\mathbf{E}_S(\mathbf{r}, \omega), \mathbf{H}_S(\mathbf{r}, \omega)] \approx \sum_m \alpha_m(\omega)\, \text{NFFT}_{\text{Re}(\tilde{\omega}_m)}[\tilde{\mathbf{E}}_m(\mathbf{r}), \tilde{\mathbf{H}}_m(\mathbf{r})], \tag{15}$$

and $\text{NFFT}_{\text{Re}(\tilde{\omega}_m)}[\tilde{\mathbf{E}}_m(\mathbf{r}), \tilde{\mathbf{H}}_m(\mathbf{r})]$ becomes a quantity that is intrinsic to the QNM and that is computed once and for all. Equation (15) is approximate; it is accurate only if the QNMs with a resonance frequency $\tilde{\omega}_m \approx \omega$ are considered in the expansion.

The build-in near-to-far-field transform of COMSOL Multiphysics is valid only for objects surrounded by uniform media (free space). Thus, we recommend that the user downloads [29] and installs the freeware RETOP [22] before using the toolbox. RETOP operates for objects on substrates or buried in stratified media and computes the radiation diagram in the superstrate and substrate. When the substrate supports guided modes or surface plasmon modes, RETOP additionally computes the in-plane radiation diagram of the guided modes.

The knowledge of the far-field radiated by a QNM is useful in practice. For instance, when an excited molecule dominantly decays into a bright QNM, an important question that arises is how the nanoresonator redirects the emitted light in the far-field or which numerical aperture is needed to collect most of the radiated light. This is central to access QNM brightness [12] and even optimize nanoresonator geometries to enhance the brightness through inverse design.

Note that the approximation made by computing the near-to-far-field transform at $\text{Re}(\widetilde{\omega}_m)$ in Eq. (15) can be removed [30].

### 4.6 Multipole expansion

Interference between multipole moments of photonic or plasmonic nanoresonators is known to lead to many exotic phenomena, like zero forward or backward scattering, non-radiating anapole responses, and Fano resonances [31,32]. A standard approach to analyze the multipolar response of a resonator consists in expanding the scattered field upon a specific excitation into spherical harmonics [33]. The multipole analysis is thus performed for a particular instance of the driving field (wavelength, incident direction, and polarization) it is not intrinsic to the resonator and does not give any insight into the resonator properties at other wavelengths.

In contrast, the *multipole* toolbox analyzes the multipolar content of QNMs at complex frequencies. It provides transparent physics since the multipolar content becomes an intrinsic property of the resonance, irrespectively of the illumination condition. This markedly changes design perspectives toward operation over a broad range of frequencies or incidence angles or both [9].

The multipolar decomposition at complex frequencies implemented in the toolbox is analogous to the decomposition performed at real frequencies [33]. When the resonator is immersed in a homogenous medium with a refractive index $n_b$, the multipolar content of a QNM is determined by expanding its electric field $\widetilde{\mathbf{E}}$ in vector spherical wave functions (VSWFs) [9]

$$\widetilde{\mathbf{E}}(\mathbf{r}) = \widetilde{k}^2 \sum_{n=1}^{\infty} \sum_{m=-n}^{n} E_{nm} \left[ \widetilde{a}_{nm} \widetilde{\mathbf{N}}_{nm}^{(3)}(\mathbf{r}, \widetilde{\omega}) + \widetilde{b}_{nm} \widetilde{\mathbf{M}}_{nm}^{(3)}(\mathbf{r}, \widetilde{\omega}) \right], \tag{16}$$

where $\widetilde{a}_{nm}$ and $\widetilde{b}_{nm}$ are electric and magnetic multipole coefficients, $\widetilde{\mathbf{N}}_{nm}^{(3)}(\mathbf{r}, \widetilde{\omega})$ and $\widetilde{\mathbf{M}}_{nm}^{(3)}(\mathbf{r}, \widetilde{\omega})$ are the outgoing VSWFs and $\widetilde{k} = \widetilde{\omega}/c$ is the complex wavevector of QNM. The *Multipole* toolbox computes the coefficients $\widetilde{a}_{nm}$ and $\widetilde{b}_{nm}$ by performing an inner product between the QNM field and the vector spherical wave functions (VSWFs) on a sphere surface circumscribing the resonator (see Fig. 8(a)). The Cartesian multipole moments of the QNM, electric dipole $\widetilde{\mathbf{p}}$, magnetic dipole $\widetilde{\mathbf{m}}$, and electric quadrupole $\widetilde{\mathbf{Q}}^e$, can then be retrieved by matching their far-field expressions with those of the VSWFs in spherical coordinates [9,31]. The relation between the Cartesian multipole moments and the VSWF coefficients can be found in Ref. [9].

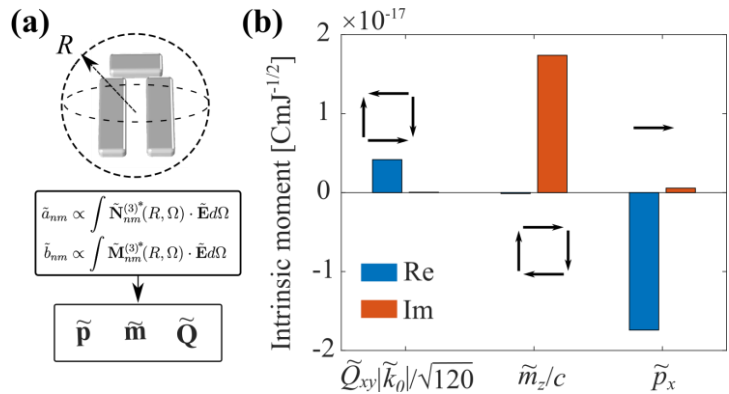

**Fig. 8. (a)** The *multipole* toolbox computes the coefficients $\widetilde{a}_{nm}$ and $\widetilde{b}_{nm}$ by performing an inner product between the QNM field and the vector spherical wave functions (VSWFs). **(b)** Multipole content of a Dolmen QNM provided by the *multipole* toolbox. The QNM is composed of electric and magnetic dipole moments, $\widetilde{p}_x$ and $\widetilde{m}_z$, which have similar amplitudes and are out of phase by $\approx 90°$ (Janus particle). The electric quadrupole component, $\widetilde{Q}_{xy}$, is real and three time smaller.

Figure 8(b) shows the example proposed in the toolbox, which may be reproduced by running the MATLAB script main_field_multipole.m. The Dolmen (same structure as Fig. 5) is designed so that $\tilde{p}_x$ and $\tilde{m}_z/c$ dominantly respond, have similar amplitudes, and are out of phase by approximately 90° (blue and red represent the real and imaginary parts) to implement a Janus response [9].

**4.7 Non-reciprocal materials**

Non-reciprocal materials are usually found in magneto-optical materials biased by an external magnetic field. They are characterized by non-symmetric permeability or permittivity tensors, $\boldsymbol{\mu} \neq \boldsymbol{\mu}^T$ or $\boldsymbol{\varepsilon} \neq \boldsymbol{\varepsilon}^T$. The orthonormalization of QNMs deserves specific attention [3,12,13,34] in this case. It requires computing a left QNM, $\tilde{\mathbf{E}}_m^{(L)}$, which can be found by solving the eigenvectors of the source-free Maxwell operator with transposed permittivity and permeability, $\boldsymbol{\varepsilon}^T$ and $\boldsymbol{\mu}^T$. Note that, for nonreciprocal systems, the left $\left[\tilde{\mathbf{E}}_m^{(L)}, \tilde{\mathbf{H}}_m^{(L)}\right]$ and right $\left[\tilde{\mathbf{E}}_m^{(R)}, \tilde{\mathbf{H}}_m^{(R)}\right]$ QNMs differ: $\tilde{\mathbf{H}}_n^{(L)} \neq \tilde{\mathbf{H}}_n^{(R)}$ and $\tilde{\mathbf{E}}_n^{(L)} \neq \tilde{\mathbf{E}}_n^{(R)}$ (the right QNMs are the eigenvectors discussed in all other sections obtained for $\boldsymbol{\varepsilon}$ and $\boldsymbol{\mu}$ tensors). However, the left and right QNMs share the same eigenvalues, $\tilde{\omega}_n^{(R)} = \tilde{\omega}_n^{(L)} \equiv \tilde{\omega}_n$.

The expression of the QNM norm $QN_n$ for non-reciprocal systems is [3,12,13]

$$QN_n = \iiint_{\Omega \cup \Omega_{\text{PML}}} \tilde{\mathbf{E}}_n^{(L)} \cdot \frac{\partial(\omega\boldsymbol{\varepsilon})}{\partial \omega} \tilde{\mathbf{E}}_n^{(R)} - \tilde{\mathbf{H}}_n^{(L)} \cdot \frac{\partial(\omega\boldsymbol{\mu})}{\partial \omega} \tilde{\mathbf{H}}_n^{(R)} dV, \qquad (17)$$

where the integral is performed in the PML layer $\Omega_{\text{PML}}$ and the physical space $\Omega$. Similarly, the mode volume reads as

$$\tilde{V}_n(\mathbf{r}, \mathbf{u}) = \left[2\varepsilon_0 \left(\tilde{\mathbf{E}}_n^{(L)} \cdot \mathbf{u}\right)\left(\tilde{\mathbf{E}}_n^{(R)} \cdot \mathbf{u}\right)/QN_n\right]^{-1}, \qquad (18)$$

$\mathbf{u}$ being the unit vector defining the polarization direction. For a derivation of these formulas and the related orthogonality condition, the reader may refer to Ref. [3] and the documentation of the *nonreciprocal_resonators* toolbox [13]. The latter presents a COMSOL model of a didactical example for computing and normalizing the non-reciprocal QNMs of a 2D yttrium-iron garnet wire in air. In this example, $\tilde{\mathbf{E}}_n^{(L)}$ and $\tilde{\mathbf{E}}_n^{(R)}$ largely differs, see Fig. 8 of [3]. Therefore, to compute $QN_n$ or $\tilde{V}_n(\mathbf{r}, \mathbf{u})$, the eigensolver should be called twice.

**4.8 Nonlinear nanophononics**

The *nonlinear* toolbox illustrates a generic example of how to perform a QNM analysis of a second-harmonic generation with localized resonances in the small-signal regime. It promotes important nonlinear concepts such as mode overlap, field enhancement, and phase matching between $\omega$ and $2\omega$ signals.

With the small-signal approximation, the pump depletion can be neglected, and the second-harmonic generation is described via two coherent processes. Under excitation by an external pump beam $\mathbf{E}_b(\mathbf{r}, \omega)$, a total field distribution $\mathbf{E}_{tot}(\mathbf{r}, \omega)$ is generated at the fundamental frequency $\omega$. It further induces a local nonlinear current inside the resonator, $\mathbf{J}^{(2)}(\mathbf{r}, 2\omega)$, which acts as the source for the second-harmonic radiation at $2\omega$.

In the *nonlinear* toolbox, the total field $\mathbf{E}_{tot}(\mathbf{r}, \omega)$ is expanded using Eq. (6) and **M 2** in Table 3. The nonlinear displacement current, $\mathbf{J}^{(2)}(\mathbf{r}, 2\omega)$, is then derived from the second-order nonlinear susceptibility $\chi^{(2)}$ and the total field $\mathbf{E}_{tot}^{(2)}(\mathbf{r}, 2\omega)$ generated at $2\omega$ is reconstructed in the QNM basis $\mathbf{E}_{tot}^{(2)}(\mathbf{r}, 2\omega) = \sum_m \alpha_m^{(2)}(2\omega) \tilde{\mathbf{E}}_m(\mathbf{r})$. The analytically known modal coefficients $\alpha_m^{(2)}(2\omega)$ are computed by performing overlap integrals between the QNMs and the nonlinear current $\mathbf{J}^{(2)}(\mathbf{r}, 2\omega)$. Finally, from the knowledge of $\mathbf{E}_{tot}^{(2)}(\mathbf{r}, 2\omega)$, the nonlinear extinction cross-section $\sigma_{ext}^{(2)}(2\omega)$, a classical figure of merit defined as the ratio between the generated power at $2\omega$ and the intensity of the incident field, is derived.

In the *nonlinear* toolbox, the second-harmonic generation is presented for an AlGaAs nanocylinder on a glass substrate [10]. The generic example may help the reader to implement other geometries or nonlinear processes. Note the recent work [35] that uses the same formalism for the third-order harmonic generation.

### 4.9 Grating analysis and band diagram of crystal

Grating spectra exhibit sharp variations of the scattered light, known as grating anomalies or resonances, which are used in many applications. Using the example of a 1D periodic grating illuminated by a plane wave at oblique incidence, the *grating* toolbox illustrates how to compute and normalize the QNMs of periodic structures, to reconstruct the scattered field in the QNM basis.

In many works [21,36,37], the QNMs are defined for a fixed in-plane (parallel to the $x$-direction in Fig. 8) Bloch-wavevector $k_p$. $k_p$-QNMs are relevant for computing band diagrams but are irrelevant for analyzing grating spectra [8].

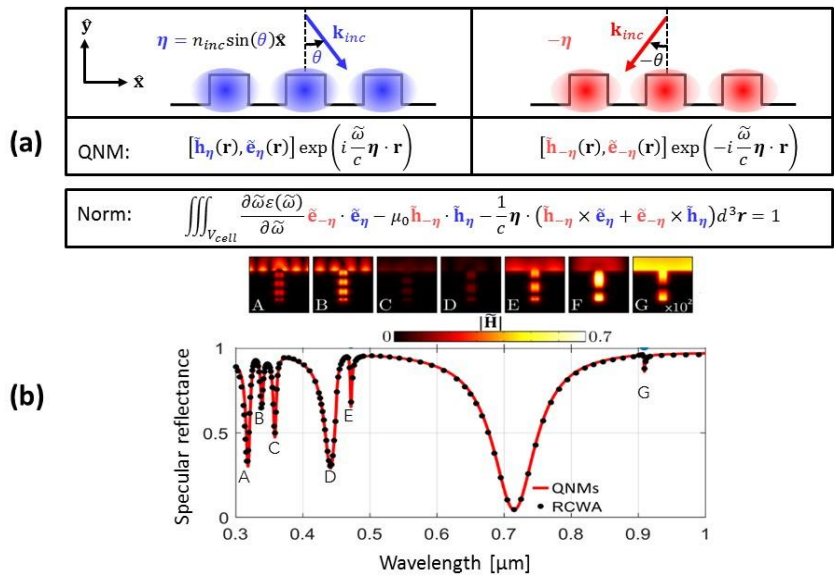

**Fig. 9**. *grating* toolbox. (**a**) The normalization of $\boldsymbol{\eta}$-QNMs with a fixed "directionality" vector $\boldsymbol{\eta}$ relies on two reciprocal QNMs with "directionality" vectors $\boldsymbol{\eta}$ and $-\boldsymbol{\eta}$. (**b**) Reconstruction of the reflectance spectrum of a gold lamellar grating illuminated by a TM-polarized plane wave ($\theta = 30°$). The reconstructed spectrum obtained with 200 modes is compared with reference data obtained with the RCWA [37]. The upper inset shows the magnetic-field moduli of the seven dominant QNMs, labeled A, B, … G. More details are found in the documentation and in [26].

The specificity of the *grating* toolbox is that it considers resonances that are effectively revealed by any experimental grating spectra measured for fixed incident angles. The resonances are QNMs obtained for a fixed angle of incidence ($\boldsymbol{\eta}$-QNMs), i.e., a fixed "directionality" vector $\boldsymbol{\eta} = n_{inc}\sin\theta\hat{\boldsymbol{x}}$. $\boldsymbol{\eta}$ is a real vector that depends on the incident angle $\theta$ and the refractive index $n_{inc}$ of the incident medium, but not on the frequency $\omega$. $\boldsymbol{k}_p$- and $\boldsymbol{\eta}$-QNMs are intrinsically different and have distinct normalizations [8]. Similarly, in a sensing experiment, the shift in resonance frequency is due to a perturbation of $\boldsymbol{\eta}$-QNMs, not $\boldsymbol{k}_p$-QNMs, except if, unlikely, one records the spectra by simultaneously tuning the angle of incidence for every frequency of measurement.

The normalization of a $\boldsymbol{\eta}$-QNM requires computing two QNMs: one with a "directionality" vector $\boldsymbol{\eta}$ and another one with another "directionality" vector $-\boldsymbol{\eta}$, see Fig.9. The $-\boldsymbol{\eta}$-QNM should be computed numerically, except if the grating has a mirror symmetry with respect to the vertical axis $y$, in which case, it can be deduced analytically from the $\boldsymbol{\eta}$-QNM.

Figure 9 illustrates one of the few reconstructions available in the literature with $\boldsymbol{\eta}$-QNMs [8].

# 5 Theoretical and numerical issues

## 5.1 The QNMPole and QNMEig solvers

**QNMPole**. Refer to Section 2.3.2 for a detailed description of the algorithm implemented in **QNMPole**.

**QNMEig**. The solver extends the existing possibilities of the eigenmode solver of the COMSOL Multiphysics RF Module. It solves a quadratic eigenvalue problem using auxiliary fields [7,38-40] to handle many material dispersions of current interest, e.g. Lorentz-Drude or $N$-pole Lorentz dispersions (Table 1). Other types of dispersion can be implemented by the user following the linearization procedure implemented in **QNMEig**. For the sake of illustration, let us consider the case of a material with a Lorentz-Drude dispersion $\varepsilon(\omega)/\varepsilon_\infty = 1 - \sum_{i=1}^{N} \frac{\omega_{p,i}^2}{\omega^2 - \omega_{0,i}^2 - i\omega\gamma_i}$. The quadratic eigenvalue formulation of the QNM reads as [7]

$$\widehat{\mathbf{K}}\boldsymbol{u} + \widetilde{\omega}\widehat{\mathbf{C}}\boldsymbol{u} + \widetilde{\omega}^2 \widehat{\mathbf{M}}\boldsymbol{u} = \mathbf{0}. \tag{19}$$

Equation (19) formally corresponds to the Lorentz oscillator model. In the finite-element-method terminology,

$$\widehat{\mathbf{K}} = \begin{bmatrix} \nabla \times \mu_0^{-1} \nabla \times & 0 & 0 & \dots \\ \varepsilon_\infty \omega_{p,1}^2 & -\omega_{0,1}^2 & 0 & \dots \\ \varepsilon \omega_{p,2}^2 & 0 & -\omega_{0,2}^2 & \dots \\ \vdots & \vdots & \vdots & \ddots \end{bmatrix}, \tag{20}$$

$$\widehat{\mathbf{C}} = \begin{bmatrix} 0 & 0 & 0 & \dots \\ 0 & -i\gamma_1 & 0 & \dots \\ 0 & 0 & -i\gamma_2 & \dots \\ \vdots & \vdots & \vdots & \ddots \end{bmatrix}, \tag{21}$$

$$\widehat{\mathbf{M}} = \begin{bmatrix} -\varepsilon_\infty & -1 & -1 & \dots \\ 0 & 1 & 0 & \dots \\ 0 & 0 & 1 & \dots \\ \vdots & \vdots & \vdots & \ddots \end{bmatrix}. \tag{22}$$

are the so-called stiffness, damping, and mass matrices, respectively. In Eq. (19), $\boldsymbol{u}$ is the augmented field vector,

$$\boldsymbol{u} = [\widetilde{\mathbf{E}} \quad \widetilde{\mathbf{P}}_1 \quad \widetilde{\mathbf{P}}_2 \quad \dots], \tag{23}$$

which includes the electric field $\widetilde{\mathbf{E}}$ and the auxiliary polarization fields $\widetilde{\mathbf{P}}_i$ ($\widetilde{\mathbf{P}}_i \equiv -\varepsilon_\infty \frac{\omega_{p,i}^2}{\widetilde{\omega}^2 - \omega_{0,i}^2 - i\widetilde{\omega}\gamma_i} \widetilde{\mathbf{E}}$).

The validity of Eq. (19) can be checked by replacing $\widetilde{\mathbf{P}}_i$ by its expression in terms of $\widetilde{\mathbf{E}}$ to obtain the Helmholtz equation. Equation (19) is fully compatible with the COMSOL eigenmode solver that solves quadratic eigenproblems with a remarkable efficiency through the so-called first companion linearization

$$\begin{bmatrix} \widehat{\mathbf{K}} & \widehat{\mathbf{C}} \\ \mathbf{0} & \mathbf{1} \end{bmatrix} \begin{bmatrix} \boldsymbol{u} \\ \boldsymbol{v} \end{bmatrix} + \widetilde{\omega} \begin{bmatrix} \mathbf{0} & \widehat{\mathbf{M}} \\ -\mathbf{1} & \mathbf{0} \end{bmatrix} \begin{bmatrix} \boldsymbol{u} \\ \boldsymbol{v} \end{bmatrix} = 0, \tag{24}$$

with $\boldsymbol{v} \equiv \widetilde{\omega}\boldsymbol{u}$.

The implementation in COMSOL Multiphysics starts by reformulating Eq. (19) into an integral form called the weak formulation. In this formulation, as suggested by its name, differential equations are no longer satisfied exactly at every point of the mesh. Instead, they are satisfied in a "weak" sense by considering an overlap integral that involves the multiplication of the equation by certain test (basis) functions. One unique advantage of the weak formulation is that the order of a differential equation can be reduced by using the method of integration by parts to improve numerical stability.

Consider the first row of Eq. (19) for example,

$$\nabla \times \mu_0^{-1} \nabla \times \tilde{\mathbf{E}} - \widetilde{\omega}^2 \varepsilon_\infty \tilde{\mathbf{E}} - \widetilde{\omega}^2 \widetilde{\mathbf{P}} = 0. \tag{25}$$

The weak formulation is obtained by introducing a test function, $\tilde{\mathbf{E}}_{\text{test}}$, and by evaluating the following overlap integral in the simulation domain $\mathbf{\Omega}$

$$\int_\Omega \tilde{\mathbf{E}}_{\text{test}}(\mathbf{r}) \cdot \nabla \times \mu_0^{-1} \nabla \times \tilde{\mathbf{E}}(\mathbf{r}) - \widetilde{\omega}^2 \varepsilon_\infty \tilde{\mathbf{E}}_{\text{test}}(\mathbf{r}) \cdot \tilde{\mathbf{E}}(\mathbf{r}) - \widetilde{\omega}^2 \tilde{\mathbf{E}}_{\text{test}}(\mathbf{r}) \cdot \widetilde{\mathbf{P}}(\mathbf{r}) \, d^3\mathbf{r} = 0. \tag{26}$$

The second-order derivative of Eq. (25) is then reduced to a first-order derivative by the technique of integration by parts,

$$\int_\Omega \nabla \times \tilde{\mathbf{E}}_{\text{test}}(\mathbf{r}) \cdot \mu_0^{-1} \nabla \times \tilde{\mathbf{E}}(\mathbf{r}) - \widetilde{\omega}^2 \varepsilon_\infty \tilde{\mathbf{E}}_{\text{test}}(\mathbf{r}) \cdot \tilde{\mathbf{E}}(\mathbf{r}) - \widetilde{\omega}^2 \tilde{\mathbf{E}}_{\text{test}}(\mathbf{r}) \cdot \widetilde{\mathbf{P}}(\mathbf{r}) \, d^3\mathbf{r} = 0, \tag{27}$$

in which we assume that the electric fields vanish on the outer boundaries of the simulation domain. The tutorials of **QNMEig** provide a step-by-step presentation of how to implement the weak formulation with the COMSOL syntax.

**5.2 QNM normalization**

Normalization enables a direct assessment of the QNM excitation strength $\alpha_m$ used for reconstruction [4]. It also enables a direct assessment one of the main figures of merit of QNM, the mode volume $\tilde{V}$ [12], which quantifies the interaction strength of the mode with point-like dipoles: $\tilde{V} \propto \tilde{\mathbf{E}}^{-2}$ ($\tilde{\mathbf{H}}^{-2}$) for dipolar electric (magnetic) interactions, $\tilde{\mathbf{E}}$ and $\tilde{\mathbf{H}}$ being the normalized fields. QNM normalization thus represents an essential step in QNM theory.

The normalization of QNMs (i.e. of open non-Hermitian systems) differs from the classical energy-based norm used for normalizing the normal modes of closed Hermitian systems. Owing to the field divergence outside the resonator, the literature on QNM normalization, including milestone publications, is quite confusing, sometimes misleading, and not free of mistakes. We refer the reader to Section 4 of the recent review [3], in which a thorough comparison of the main normalization methods used during the last decade has enabled to highlight mistakes and fully clarify the respective domain of validity of every method. For newcomers, let us mention that, albeit more complicated mathematically, the QNM norm is as simple as the normal-mode norm to implement numerically.

Two markedly different normalization methods are used in **MAN**. Albeit very different in form, they are mathematically equivalent [3,5] and, they are both very easy to implement numerically; they lead to quasi-identical (if the same mesh is used for their computation, more than 11 common digits are obtained) numerical values (Fig. 6 in Ref. [3]).

The **QNMEig** and **QNMPole** solvers implemented in **MAN** are dedicated to reciprocal materials for which the right and left QNMs (the left eigenvectors are solutions of the source-free transposed Maxwell operator, see Appendix C in [3]) are identical. For periodic structures (see Section 4.9) or resonators with non-reciprocal materials (see Section 4.7), the right and left eigenvectors are different, and the normalization thus requires performing two independent computations, one for the right QNMs and the other for the left ones. The *nonreciprocal* and *grating* toolboxes are dedicated to these two special cases.

**QNMEig norm.** The **QNMEig** solver uses the so-called PML-norm [4]. As the name of the method suggests, this norm requires that perfectly matched layers (PMLs) be used to satisfy the outgoing wave conditions at the boundary of the computational domain. In practice, the PML-norm, $QN = \iiint_{\Omega \cup \Omega_{PML}} \tilde{\mathbf{E}}_m \cdot \frac{\partial \omega \varepsilon}{\partial \omega} \tilde{\mathbf{E}}_m - \tilde{\mathbf{H}}_m \cdot \frac{\partial \omega \mu}{\partial \omega} \tilde{\mathbf{H}}_m \, dV$, is easily computed as a volume integral over the entire computational domain composed of the inner physical domain $\Omega$ and the surrounding PML domain $\Omega_{PML}$. Note that a PML is formally equivalent to a complex coordinate stretching [41,42]. The PML domain $\Omega_{PML}$, therefore denotes a volume in complex space. In practice, in **MAN**, the volume integral in the $\Omega_{PML}$ domain is simply obtained by multiplying the real-space integrand by a Jacobian to accommodate for the change of coordinates : $\iiint_{\Omega_{PML}} \ldots dV = \iiint_{V_{PML}} \ldots \|\bar{\bar{J}}\| dV$ with $\|\bar{\bar{J}}\|$ the Jacobian, and $V_{PML}$ the volume of the PML layer mapped onto the real coordinates.

**QNMPole** norm. The normalization adopted by the **QNMPole** solver does not rely on the calculation of any volume integrals and does not require PMLs (Section 2.3.2). It has been introduced in Ref. [5] and is referred to as the pole-response norm in the literature [3]. It relies on the fact that, for complex frequencies $\omega$ close to the QNM eigenfrequency $\widetilde{\omega}_m$, the scattered field is proportional to the normalized QNM field with a known excitation coefficient: $\lim_{\omega \to \widetilde{\omega}_m}(\omega - \widetilde{\omega}_m)\mathbf{E}_S(\mathbf{r}, \omega) = \lim_{\omega \to \widetilde{\omega}_m}(\omega - \widetilde{\omega}_m)\alpha_m(\widetilde{\omega}_m)\widetilde{\mathbf{E}}_m(\mathbf{r})$. As mentioned earlier, there is an infinity of expressions for $\alpha_m(\widetilde{\omega}_m)$; however, they all tend towards the same asymptotic expression $\alpha_m(\widetilde{\omega}_m)$ as $\omega \to \widetilde{\omega}_m$ and thus the normalization is unique. The pole-response method is the most general method for QNM normalization: it can be implemented with any frequency-domain Maxwell solvers and can be used for any geometries, including those that cannot accommodate PML boundary conditions, such as photonic-crystal cavities that leak into semi-infinite periodic waveguides [43].

### 5.3 Completeness of QNM expansions

Completeness refers to the possibility of rigorously reconstructing the field $[\mathbf{E}_S(\mathbf{r}, \omega), \mathbf{H}_S(\mathbf{r}, \omega)]$ scattered by the resonator in the QNM basis, at least for $\mathbf{r}$ belonging to a compact subspace of $\mathbb{R}^3$ [44-46]. We refer the reader to a recent review [3].

Disappointingly, completeness of expansions involving only true QNMs is rarely met. These expansions are complete *only inside the resonators* ($\mathbf{r} \in V_{res}$) and *for resonators in free space* (without substrate), owing to the existence of branch cuts for geometries with a substrate. As explained in [3], there might be hope for future theoretical improvements, but so far, the literature suggests only two possibilities to move towards realistic cases with resonators deposited on substrates, embedded in thin films, or coupled to waveguides.

The first possibility relies on complementing a restricted finite subset of QNMs with another contribution that might be called a non-resonant contribution. Two approaches to compute the non-resonant contribution are available. The first approach, the so-called Riesz projection method [47], relies on computing the non-resonant contribution using a finite-length contour in the complex frequency plane. Alternatively, the non-resonant contribution may also be computed by directly solving Maxwell equations at a few real frequencies, and since it is expected to gently vary with the frequency, it can be easily interpolated [11]. This second approach is implemented in the *interpolation* toolbox (Section 4.3).

The second possibility, which can be implemented with the **QNMEig** solver approach and the *reconstruction* toolbox, relies on a regularization procedure, initially launched in quantum mechanics, which transforms the open space into a regularized Hilbert space using a complex coordinate transform implemented with perfectly-matched layers [3,4,7,8,21,26]. The spectrum of the regularized operator incorporates true QNMs and numerical modes, which obey the same orthogonality product and, together, admittedly form a complete basis of the regularized ('PMLized') Hilbert space that features square-integrable vectors. A more in-depth discussion can be found in Section 3.4.3 in [3].

State-of-the-art reconstructions of the scattered field by increasing the total number of QNMs and numerical modes retained in the expansion have been obtained with the **QNMEig** solver [7]. The results are quite convincing since convergence is numerically achieved over a broad range of frequencies for complicated geometries: resonators with dispersive materials, on dielectric substrates with guiding layers (with branch cuts), on metallic substrates (with branch cuts and accumulation points). In [26], the authors consider simpler geometries, namely 2D cylinders and spheres in free space. They have numerically verified that expansions with QNMs and numerical modes are complete both inside and outside the resonator. They have also verified that expansions based on QNMs only are incomplete everywhere, even inside the resonator, owing to the branch-cut of the Green's function in 2D, see Fig. 2 in [26] for more details. The convergence has also been numerically demonstrated in the temporal domain [7]. Another example of converged results is found in Fig. 11 in [8], in which a metal grating with many branch cuts (corresponding to the passing-off of diffraction orders) is analyzed.

### 5.4 Different methods for reconstruction

Table 3 presents $2 \times 4$ methods for reconstructing the total field $\mathbf{E}_{tot}$. This section qualitatively explains the origin for the existence of various methods.

As the reader may have noted, in Table 3, the analytical expressions for the reconstruction depend on the formula, Eq. (5) or Eq. (6), used to compute the excitation coefficient $\alpha_m$. This dependence results from the projection of different state vectors onto the QNM basis [26]. The projected state vectors are $[\mathbf{E}_S, \mathbf{H}_S, \mathbf{P}_S, \mathbf{J}_S]^T$ for Eq. (5) and $[\mathbf{E}_S, \mathbf{H}_S, \mathbf{P}_{tot}, \mathbf{J}_{tot}]^T$ for Eq. (6). Therefore, when adopting Eq. (5), the scattered auxiliary fields can be decomposed into a sum of QNMs, $[\mathbf{P}_S, \mathbf{J}_S] = (\varepsilon(\omega) - \varepsilon_\infty)[\mathbf{E}_S, i\omega \mathbf{E}_S] = \sum_m \alpha_m [\widetilde{\mathbf{P}}_m, \widetilde{\mathbf{J}}_m]$, whereas, when adopting Eq. (6), the same formula holds for the total auxiliary fields, $[\mathbf{P}_{tot}, \mathbf{J}_{tot}] = (\varepsilon(\omega) - \varepsilon_\infty)[\mathbf{E}_{tot}, i\omega \mathbf{E}_{tot}] = \sum_m \alpha_m [\widetilde{\mathbf{P}}_m, \widetilde{\mathbf{J}}_m]$.

Let us now explain how the method **M 4** is obtained. We start by reconstructing the total magnetic field with the simplest formula, $\mathbf{H}_{tot} = \sum_m \alpha_m \widetilde{\mathbf{H}}_m + \mathbf{H}_b$. Then we use the Maxwell equation, $\mathbf{E}_{tot} = [i\omega\varepsilon(\omega)]^{-1} \nabla \times \mathbf{H}_{tot}$, to deduce the $\mathbf{E}_{tot}$ (method **M 4** in Table 3). Indeed, when all the QNMs are retained in the expansion, the $\mathbf{E}_{tot}$ expressions of methods **M 4** and **M 1** ($\mathbf{E}_{tot} = \sum_m \alpha_m \widetilde{\mathbf{E}}_m + \mathbf{E}_b$) are mathematically equivalent; however, when truncating the series, they differ. Likewise, for dispersive resonators, $\mathbf{E}_{tot}$ can also be deduced from the reconstructed auxiliary fields $\mathbf{P}_{tot}$ and $\mathbf{J}_{tot}$, therein leading to methods **M 2** and **M 3**.

It is natural to wonder which reconstruction method provides the most accurate prediction by retaining the smallest number of QNMs in the expansion. According to our experience, there is no general answer to that question. When reconstructing the fields for plasmonic resonators [9] and high-index dielectric Mie nanoresonators [10], we have observed that the accuracies achieved with methods **M 2**, **M 3**, **M 4** are comparable when considering dozens of QNMs. Methods **M 3** and **M 4** may have a slightly slower convergence for high-index nanoresonators, possibly because the contribution of numerous high-frequency modes is boosted with the prefactor $\widetilde{\omega}_m/\omega$.

On the other hand, we have observed that the simplest and most widespread reconstruction method **M 1** ($[\mathbf{E}_{tot}, \mathbf{H}_{tot}] = \sum_m \alpha_m [\widetilde{\mathbf{E}}_m, \widetilde{\mathbf{H}}_m] + [\mathbf{E}_b, \mathbf{H}_b]$) is rarely capable of providing very accurate reconstructions (an example is provided in Fig. 5), except for some simple cases for which the optical response can be fully described with 1 or 2 dominant QNMs [4,5]. The reason can be intuitively understood. Consider the reconstruction of the total electric field of a non-dispersive resonator and imagine that the resonator is illuminated by a plane wave. The continuity of the normal component of the electric displacement field across the boundary of the resonator $\mathbf{n} \cdot \mathbf{D}_{tot+} = \mathbf{n} \cdot \mathbf{D}_{tot-}$ has to be fulfilled. However, since the background electric field $\mathbf{E}_b$ is a continuous function across the boundary of the resonator, the reconstruction with 'transverse' QNMs satisfying $\mathbf{n} \cdot \widetilde{\mathbf{D}}_{m+} = \mathbf{n} \cdot \widetilde{\mathbf{D}}_{m-}$ naturally results in a discontinuity $\mathbf{n} \cdot (\varepsilon_+ \mathbf{E}_{b+} - \varepsilon_- \mathbf{E}_{b-})$ of the normal component of the electric displacement, which can only be compensated by the contribution of a large set of longitudinal QNMs satisfying $\mathbf{n} \cdot \widetilde{\mathbf{D}}_{m+} \neq \mathbf{n} \cdot \widetilde{\mathbf{D}}_{m-}$. These longitudinal QNMs satisfy the source-free Maxwell equations at zero frequency, $\nabla \times \widetilde{\mathbf{E}}_m = 0$, and are called electric static QNMs [25,48] in the literature. We have not yet understood for which geometry their contribution is significant or negligible. However, if their contribution cannot be neglected, they may require computing a large number of numerical electrostatic-like modes with complex frequencies close to the zero frequency. On a few occasions, we have also computed them with the electrostatic solver of COMSOL [49]. Note that the contribution of the electrostatic-like modes is a smooth function of the frequency in the visible and is easily computed with the *interpolation* toolbox.

### 5.5 Symmetry

**Mirror Symmetry.** Many optical resonators exhibit mirror symmetry, i.e., the resonator can be superimposed with itself by flipping it with respect to one or several planes. It is beneficial to implement perfect electric/magnetic conductors (PEC/PMC) at the symmetry planes and only mesh one-half, one-quarter, or even one-eighth of the systems, to speed up the QNMs computation by approximately one order of magnitude in 3D. Depending on whether the implemented boundary condition at the symmetry plane is PMC or PEC, QNMs with the specific mirror symmetry will be computed. For example, for a PEC in the $x$-$y$ plane, only the QNM with $\left[\tilde{E}_x(\mathbf{r}), \tilde{E}_y(\mathbf{r}), \tilde{E}_z(\mathbf{r})\right] = \left[-\tilde{E}_x(\mathbf{r}'), -\tilde{E}_y(\mathbf{r}'), \tilde{E}_z(\mathbf{r}')\right]$, with $\mathbf{r} = [x, y, z]$ and $\mathbf{r}' = [x, y, -z]$, will be computed in the upper half-space ($z > 0$). **MAN** includes built-in functions to automatically derive the QNM fields in the unmeshed space, in such case the lower half-space ($z < 0$).

**Axisymmetry.** 3D axisymmetric optical nanoresonators are widely experienced in nanophotonics. Typical examples include whispering gallery resonators, spherical dimers, nanodisks, and so on. The QNM computation for these geometries can be simplified to a 2D problem by implementing axial symmetry. The average time for computing one QNM can be reduced by a few orders of magnitude, and numerical stability can be greatly enhanced.

In the cylindrical coordinate system $\{\rho, \phi, z\}$, the axisymmetric QNM fields can be expressed as

$$\widetilde{\mathbf{E}}_m(\mathbf{r}) = \tilde{\mathbf{e}}_m(\rho, z) \exp(im\phi), \text{ and } \widetilde{\mathbf{H}}_m(\mathbf{r}) = \tilde{\mathbf{h}}_m(\rho, z) \exp(im\phi), \tag{28}$$

where $m = 0, \pm 1, \pm 2, \cdots$ denotes the azimuthal order of the mode. The azimuthal dependence of the fields $\exp(im\phi)$ being explicit, the determination of $\tilde{\mathbf{e}}_m(\rho, z)$ and $\tilde{\mathbf{h}}_m(\rho, z)$ corresponds to a 2D problem since only two variables, $\rho$ and $z$, are involved. The COMSOL model QNMEig_axi_NPoM.mph illustrates how to implement the axial symmetry for metal nanocylinders on a metal substrate (NPoM). It may also help the user to develop their own models.

For reciprocal systems, $\boldsymbol{\mu} = \boldsymbol{\mu}^T$ and $\boldsymbol{\varepsilon} = \boldsymbol{\varepsilon}^T$ and the normalization of axisymmetric QNMs involves $\{\tilde{\mathbf{e}}_m(\rho, z), \tilde{\mathbf{h}}_m(\rho, z)\}$ and $\{\tilde{\mathbf{e}}_{-m}(\rho, z), \tilde{\mathbf{h}}_{-m}(\rho, z)\}$

$$QN = 2\pi \iint_{\Omega \cup \Omega_{\text{PML}}} \rho \left( \tilde{\mathbf{e}}_{-m} \cdot \frac{\partial(\omega \boldsymbol{\varepsilon})}{\partial \omega} \tilde{\mathbf{e}}_m - \tilde{\mathbf{h}}_{-m} \cdot \frac{\partial(\omega \boldsymbol{\mu})}{\partial \omega} \tilde{\mathbf{h}}_m \right) d\rho dz. \tag{29}$$

However since $\{\tilde{\mathbf{e}}_m(\rho, z), \tilde{\mathbf{h}}_m(\rho, z)\}$ can be directly deduced from $\{\tilde{\mathbf{e}}_{-m}(\rho, z), \tilde{\mathbf{h}}_{-m}(\rho, z)\}$, $\tilde{\mathbf{e}}_m \cdot \hat{\rho} = \tilde{\mathbf{e}}_{-m} \cdot \hat{\rho}$, $\tilde{\mathbf{e}}_m \cdot \hat{z} = \tilde{\mathbf{e}}_{-m} \cdot \hat{z}$, $\tilde{\mathbf{e}}_m \cdot \hat{\phi} = -\tilde{\mathbf{e}}_{-m} \cdot \hat{\phi}$, $\tilde{\mathbf{h}}_m \cdot \hat{\rho} = -\tilde{\mathbf{h}}_{-m} \cdot \hat{\rho}$, $\tilde{\mathbf{h}}_m \cdot \hat{z} = -\tilde{\mathbf{h}}_{-m} \cdot \hat{z}$, and $\tilde{\mathbf{h}}_m \cdot \hat{\phi} = \tilde{\mathbf{h}}_{-m} \cdot \hat{\phi}$, $QN$ can be calculated with a single QNM computation.

## 6 Conclusion

We have introduced **MAN**, an open-source MATLAB program to model light scattering by electromagnetic resonators based on quasinormal-mode expansions. The manuscript provides an overview of the possibilities offered by the software and references the key formulas implemented in the program. **MAN** also includes a comprehensive suite of self-contained geometry examples to illustrate the program's capabilities for realistic calculations and various emblematic resonators of modern photonics. We hope this program will be useful for many current geometries studied for applications at high (optical and near-IR waves) or low (RF waves) frequencies, and we welcome contributions to extend the program use cases.

**MAN** gathers two solvers, **QNMPole** and **QNMEig**, which have already acquired a good reputation as can be seen from the number of citations of the referent publications [4,5,7] or software downloads [29]. **QNMEig** provides a comprehensive interface to the commercial software COMSOL Multiphysics; **QNMPole** can be used with any software capable of solving Maxwell equations in the frequency domain. The solvers are completely different, and together, they cover a broad scope of applications, as illustrated by the increasing number of toolboxes incorporated in the software.

The present version achieves a major shift in the development of the software. We conclude below with a summary of the main new features:

- Unified use of the **QNMEig** and **QNMPole** solvers,
- User-friendly *TUTORIALS* scripts that provide a step-by-step illustration of how to compute, normalize, sort, and visualize QNMs with both QNM solvers,
- A build-in function to export QNM fields in any user-defined coordinates,
- A set of built-in functions that provide direct access to important QNM figures of merit, e.g., mode volume, brightness, radiation diagram, multipolar content,
- A dedicated toolbox for reconstructing the scattered field in the QNM basis,
- An extended set of COMSOL models encompassing a great variety of nanophotonic devices,
- Additional built-in material dielectric functions.

Overall, we expect that the new version is more didactical and will better assist the user to develop their own applications.

## Acknowledgment

The authors acknowledge the help of Marc Duruflé (INRIA Bordeaux), Rémi Faggiani (Greenerwave), Jianji Yang (Meta/FRL), Alexandre Gras, Carlo Gigli (EPFL), Qiang Bai, Mondher Besbes, Jean-Paul Hugonin and Christophe Sauvan.## Data Availability

Data underlying the results presented in this paper are not publicly available at this time but may be obtained from the authors upon reasonable request.

## Declaration of Competing Interest

The authors declare that they have no known competing financial interests or personal relationships that could have appeared to influence the work reported in this paper.

The present version achieves a major shift in the development of the software. We conclude below with a summary of the main new features:

- Unified use of the **QNMEig** and **QNMPole** solvers,
- User-friendly *TUTORIALS* scripts that provide a step-by-step illustration of how to compute, normalize, sort, and visualize QNMs with both QNM solvers,
- A build-in function to export QNM fields in any user-defined coordinates,
- A set of built-in functions that provide direct access to important QNM figures of merit, e.g., mode volume, brightness, radiation diagram, multipolar content,
- A dedicated toolbox for reconstructing the scattered field in the QNM basis,
- An extended set of COMSOL models encompassing a great variety of nanophotonic devices,
- Additional built-in material dielectric functions.

Overall, we expect that the new version is more didactical and will better assist the user to develop their own applications.

## Acknowledgment

The authors acknowledge the help of Marc Duruflé (INRIA Bordeaux), Rémi Faggiani (Greenerwave), Jianji Yang (Meta/FRL), Alexandre Gras, Carlo Gigli (EPFL), Qiang Bai, Mondher Besbes, Jean-Paul Hugonin and Christophe Sauvan.

## Data Availability

Data underlying the results presented in this paper are not publicly available at this time but may be obtained from the authors upon reasonable request.

## Declaration of Competing Interest

The authors declare that they have no known competing financial interests or personal relationships that could have appeared to influence the work reported in this paper.

## References

1. P. Lalanne, W. Yan, V. Kevin, C. Sauvan, J.-P. Hugonin, Laser Photon. Rev. 12 (2018) 1700113.
2. P. T. Kristensen, K. Herrmann, F. Intravaia, K. Busch, Adv. Opt. Photonics 12, (2020) 612.
3. C. Sauvan, T. Wu, R. Zarouf, E. A. Muljarov, P. Lalanne, Opt. Express 30 (2022) 6846.
4. C. Sauvan, J.-P. Hugonin, I.S. Maksymov, P. Lalanne, Phys. Rev. Lett. 110 (2013) 237401.
5. Q. Bai, M. Perrin, C. Sauvan, J.-P. Hugonin, P. Lalanne, Opt. Express 21 (2013) 27371.
6. R. Faggiani, A. Losquin, J. Yang, E. Mårsell, A. Mikkelsen, P. Lalanne, ACS Photonics 4 (2017) 897.
7. W. Yan, R. Faggiani, P. Lalanne, Phys. Rev. B 97 (2018) 205422.
8. A. Gras, W. Yan, P. Lalanne, Opt. Lett. 44 (2019) 3494.
9. T. Wu, A. Baron, P. Lalanne, K. Vynck, Phys. Rev. A 101 (2020) 011803(R).
10. C. Gigli, T. Wu, G. Marino, A. Borne, G. Leo, P. Lalanne, ACS Photonics 7 (2020) 1197.
11. T. Wu, D. Arrivault, M. Duruflé, A. Gras, F. Binkowski, S. Burger, W. Yan, P. Lalanne, J. Opt. Soc. Am. A 38 (2021) 1224.
12. T. Wu, M. Gurioli, P. Lalanne, ACS Photonics 8 (2021) 1522.
13. T. Wu, P. Lalanne, arXiv:2106.05502 (2021).
14. https://www.ansys.com/products/electronics/ansys-hfss
15. https://www.3ds.com/products-services/simulia/products/cst-studio-suite/
16. A.F. Oskooi, D. Roundy, M. Ibanescu, P. Bermel, J.D. Joannopoulos, S.G. Johnson, Comput. Phys. Comm. 181 (2010) 687.
17. https://www.ansys.com/products/photonics/fdtd